\documentclass[prd,twocolumn,showpacs]{revtex4}
\usepackage{epsfig}
\usepackage{graphicx}
\usepackage{amsmath}
\usepackage{color}

\newcommand{\comment}[1]{}

\newcommand\lsim{\mathrel{\rlap{\lower4pt\hbox{\hskip1pt$\sim$}}
        \raise1pt\hbox{$<$}}}
\newcommand\gsim{\mathrel{\rlap{\lower4pt\hbox{\hskip1pt$\sim$}}
        \raise1pt\hbox{$>$}}}

\newcommand\subla{sub-Ly$\alpha$}
\newcommand\supla{super-Ly$\alpha$}
\newcommand\Subla{Sub-Ly$\alpha$}
\newcommand\Supla{Super-Ly$\alpha$}

\newcommand{\HI}{H{\sc ~i}}
\newcommand{\HII}{H{\sc ~ii}}
\newcommand{\HeI}{He{\sc ~i}}
\newcommand{\HeII}{He{\sc ~ii}}

\begin{document}

\title{Two-photon transitions in primordial hydrogen recombination}

\author{Christopher M. Hirata}
\email{chirata@tapir.caltech.edu}
\affiliation{Caltech M/C 130-33, Pasadena, California 91125, USA}

\date{March 5, 2008}

\begin{abstract}
The subject of cosmological hydrogen recombination has received much attention recently because of its importance to predictions for and cosmological 
constraints from CMB observations.  While the central role of the two-photon decay $2s\rightarrow 1s$ has been recognized for many decades, 
high-precision calculations require us to consider two-photon decays from the higher states $ns, nd\rightarrow 1s$ ($n\ge 3$).  Simple attempts to 
include these processes in recombination calculations with an effective two-photon decay coefficient analogous to the $2s$ decay coefficient 
$\Lambda_{2s}=8.22\,$s$^{-1}$ have suffered from physical problems associated with the existence of kinematically allowed sequences of one-photon 
decays, e.g. $3d\rightarrow 2p\rightarrow 1s$, that technically also produce two photons.  These correspond to resonances in the two-photon spectrum 
that are optically thick to two-photon absorption, necessitating a radiative transfer calculation.
We derive the appropriate equations, develop a numerical code to solve them, and verify the results by finding agreement with analytic 
approximations to the radiative 
transfer equation.  The related processes of Raman scattering and two-photon recombination are included using similar machinery.  
Our results show that early in recombination the two-photon decays act to speed up recombination, reducing the free electron abundance by 1.3\% 
relative to the standard calculation at $z=1300$.  However we find that some photons between Ly$\alpha$ and Ly$\beta$ are produced, mainly by 
$3d\rightarrow 1s$ two-photon decay and $2s\rightarrow 1s$ Raman scattering.  At later times these photons redshift down to Ly$\alpha$, excite hydrogen 
atoms, and act to slow recombination.  Thus the free electron abundance is increased by 1.3\% relative to the standard calculation at $z=900$.  Our 
calculation involves a very different physical argument than the recent studies of Wong \& Scott and Chluba \& Sunyaev, and produces a much larger 
effect on the ionization history.  The implied correction to the CMB power spectrum is neligible for the recently released {\slshape WMAP} and ACBAR 
data, but at Fisher matrix 
level will be $7\sigma$ for {\slshape Planck}.
\end{abstract}


\maketitle

\section{Introduction}

The anisotropies of the cosmic microwave background (CMB) are one of the most important tools for cosmology.  The power spectrum from the {\slshape 
Wilkinson Microwave Anisotropy Probe} (WMAP) satellite has enabled precision measurement of the composition of the high-redshift universe (the 
baryon/photon and dark matter/baryon ratios), the distance to the surface of last scattering, and the primordial power spectrum (including its spectral 
index $n_s$) \cite{2007ApJS..170..377S, 2008arXiv0803.0547K}.  The first two of these have proven essential to studies of dark energy, e.g. by breaking 
the approximate 
$\Omega_m-w$ degeneracy in low-redshift supernova data, while the mesurements of $n_s$ have begun to rule out interesting classes of inflationary 
models.  This trend can be expected to continue with the upcoming {\slshape Planck} satellite and several ground- and balloon-based experiments to 
measure small-scale temperature anisotropies and CMB polarization.  In particular, {\slshape Planck} should be able to map the entire CMB sky with 
signal-to-noise ratio greater than 1 down to scales of $l\sim 1600$, which in principle enables measurement of $n_s$ with an uncertainty of $\sim 
0.003$ \cite{2002PhRvD..65b3003H}.

While the raw sensitivity of the future CMB projects is spectacular, there are significant challenges for both experiment and theory.  It is well-known 
that these projects require removal of instrumental signatures, foregrounds, and secondary anisotropies, and highly accurate beam maps.  In contrast, 
the theory of primary anisotropies is generally regarded as simple: it is linear perturbation theory whose development was completed more than a 
decade ago, and which can now be solved by public computer codes that agree to within 0.1\% \cite{2003PhRvD..68h3507S}.  However, the linear 
perturbation theory calculation requires knowledge of the free electron density (and hence Thomson opacity) in the unpertubed Universe.  This free 
electron density is determined by the complicated non-equilibrium physics of recombination, as first noted forty years ago by Peebles 
\cite{1968ApJ...153....1P} and Zel'dovich et~al. \cite{1968ZhETF..55..278Z}.  For this reason, a great deal of effort in the 1990s was aimed at solving 
cosmic recombination.  This culminated in the {\sc recfast} code by Seager et~al. \cite{1999ApJ...523L...1S,2000ApJS..128..407S}, which with some 
recent improvements \cite{2008MNRAS.386.1023W} is used to compute the recombination history in most of today's CMB prediction codes. The Seager et~al. 
analysis is based on the ``multi-level atom'' (MLA) method, in which one writes down the occupation probabilities for each level of each atom or ion, 
and then constructs a set of evolution equations that includes radiative transition rates, photoionization and radiative recombination rates, and 
collision terms.  A separate allowance is made for the two-photon transition \HI\ $2s\rightarrow 1s$, and one must also include an equation for the 
matter temperature (including all heating and cooling terms) since at late times this is not in equilibrium with the CMB temperature.  The \HI\ and 
\HeII\ Lyman-series lines, and the \HeI\ $n^1P^o_1-1^1S_0$ lines, become optically thick and are treated using the Sobolev \cite{1960mes..book.....S} 
approximation.

In the past several years, there has been a resurgence of interest in the recombination problem, driven primarily by the new CMB experiments 
\cite{2006MNRAS.373..561L, 2008MNRAS.386.1023W} and also the possibility of measuring the spectral distortion 
\cite{2004AstL...30..509D, 2004AstL...30..657D, 2006MNRAS.367.1666W, 
2006MNRAS.371.1939R, 2006A&A...458L..29C, 2008A&A...478L..27C, 2008arXiv0802.0772S}.  Several groups have added additional physics to the Seager et~al. 
MLA.  This includes two-photon transitions from higher excited states of \HI\ ($ns,nd\rightarrow 1s$) \cite{2005AstL...31..359D, 2007MNRAS.375.1441W, 
2008A&A...480..629C} and \HeI\ \cite{2007astro.ph..2144H}; stimulated two-photon decays and two-photon absorption \cite{2006A&A...446...39C, 
2006AstL...32..795K}; separate consideration of each $nl$ sublevel of \HI\ instead of assuming a statistical distribution of $l$ 
\cite{2007MNRAS.374.1310C}; semiforbidden and forbidden transitions in \HeI\ \cite{2007MNRAS.375.1441W, 2007astro.ph..2143S, 2007astro.ph..2145S}; 
non-Sobolev behavior in lines with significant continuum opacity and partial redistribution \cite{2007astro.ph..2143S, 2007MNRAS.378L..39K}; and recoil 
of \HI\ atoms in Ly$\alpha$ scattering \cite{2008arXiv0801.3347G}.

This paper is concerned with the two-photon transitions during the \HII$\rightarrow$\HI\ recombination, with special attention given to the 
$ns,nd\rightarrow 1s$ decays with $n\ge 3$. The physics of the two-photon process has been well-understood since the work of G\"oppert-Mayer 
\cite{1931AnP...401..273G}, and successively more accurate computations of the $2s$ lifetime have been made in the following years 
\cite{1940ApJ....91..215B, 1951ApJ...114..407S, 2005JPhB...38..265L}.  The importance of the $2s\rightarrow 1s$ decay for cosmological recombination 
was recognized in the 1960s \cite{1968ApJ...153....1P, 1968ZhETF..55..278Z} and it is typically included in MLA codes by adding a $2s\rightarrow 1s$ 
decay with a decay rate $\Lambda_{2s}=8.22\,$s$^{-1}$ and a thermal two-photon absorption rate set by the principle of detailed balance. However, 
cosmologists paid little attention to the two-photon decays from other states until the recent work of Dubrovich \& Grachev \cite{2005AstL...31..359D} 
(hereafter DG05).  Despite the major recent advances in this subject \cite{2007MNRAS.375.1441W, 2008A&A...480..629C}, we still lack a fully 
self-consistent treatment of the problem.

DG05 attempted to extend the MLA to two-photon transitions from highly excited ($n\ge 3$) levels by defining analogous two-photon rates 
$\Lambda_{ns,nd}$.  Unfortunately they discovered that this does not work: the matrix element has poles corresponding to the ``1+1'' photon decays of 
the form $ns,nd\rightarrow Np\rightarrow 1s$ with $1<N<n$.  This results in a very fast decay rate for the $n\ge 3$ levels, with most of the decays 
producing a photon in a Lyman-series resonance.  However since the photons emitted in resonance lines have already been counted in the one-photon 
treatment, DG05 recognized that to avoid double-counting, they should somehow remove the resonance from their two-photon rates.  They did this by 
keeping only one pole in the matrix element, that corresponding to the $np$ intermediate state.  This led them to very fast decay rates that scale as 
$\Lambda_{ns,nd}\propto n$ for high $n$ and a several percent acceleration of hydrogen recombination. Wong \& Scott \cite{2007MNRAS.375.1441W} 
(hereafter WS07) compared the DG05 result for $n=3$ to calcuations \cite{1986PhRvA..33.1677C, 1988PhRvA..38.2189F} that included all non-resonant 
poles; they found a lower rate than DG05 and re-scaled DG05's $\Lambda_{ns,nd}$ values appropriately.  This resulted in a maximum change of 0.4\%\ in 
the electron abundance.  Chluba \& Sunyaev \cite{2008A&A...480..629C} (hereafter CS08) argued that one should cut off the two-photon decay rate not by 
selecting poles but based on the physical criterion of whether one of the photons is immediately re-absorbed in a Lyman-series line.  They computed 
full two-photon spectra and implemented a cutoff based on their calculation of absorption in the red damping wing of Ly$\alpha$.  This procedure has 
the distinct advantage of being convergent as $n_{\rm max}\rightarrow\infty$.  The results do depend somewhat on the nature of the cutoff 
($x_{\Gamma,c}$ in CS08 notation) but generally give changes in the electron abundance of several tenths of a percent.

In this paper, we write down the full system of equations for two-photon transitions, including emission, absorption, and Raman scattering.  We then 
take 
two approaches to solving these equations.  We first present a numerical approach in which the two-photon continuum is discretized and turned into an 
effective MLA with virtual levels.  We then present an analytic approach in which the two-photon transitions are approximated as effective corrections 
to the Ly$\alpha$ and Ly$\beta$ decay rates and added into the standard MLA with no virtual levels.  The first approach, like most fully numerical 
approaches, has the advantage of solving the two-photon equations with accuracy limited only by step sizes (in both frequency and time!) and machine 
precision.  Unfortunately, the results simply come out of the computer and are difficult to interpret.  The analytic approach allows one to 
understand on paper the cancellations that cause the two-photon effect to be finite and independent of cutoffs.  It also serves as a check on the fully 
numerical result.  In the future it may also have some value in ``fast'' recombination calculations that could be embedded in Markov chains for 
cosmological parameter estimation.

This paper is {\em not} intended to be a complete solution of the hydrogen recombination problem.  A number of effects that are not considered here 
such as the very high-$n$ levels, Ly$\alpha$ diffusion and recoil, and feedback from the helium spectral distortion \cite{1989ApJ...338..594K, 
1990ApJ...353...21K, 1994ApJ...427..603R, 2006MNRAS.367.1666W, 2007MNRAS.374.1310C,
2008arXiv0801.3347G, 2008arXiv0802.0772S}, may also be important if sub-percent accuracy is desired.  Rather, the purpose of this paper is to clear up 
the physical picture 
of the 
two-photon transitions and estimate their effect on recombination, as a first step toward a treatment that will ultimately include these other 
processes as well.

This paper is organized as follows: in Section~\ref{ss:mla} we review the multi-level atom method that is traditionally used in recombination 
studies and the sometimes-used steady-state approximation.  In Section~\ref{sec:b3} we describe how to graft two-photon transitions onto this 
framework, the problems that have occurred in past efforts to define a two-photon decay rate from highly excited states, and the resolution of these 
problems.  We describe the numerical method for solving the problem and its results in Section~\ref{sec:numeric}, and an analytic approximation in 
Section~\ref{sec:semi}.  The effect on the CMB power spectrum is assessed in Section~\ref{sec:cmb}.
We conclude in Section~\ref{sec:conc}.  The appendices cover several technical details: Appendix~\ref{app:matter} considers 
the early-time matter temperature evolution, and Appendix~\ref{app:rates} computes the two-photon decay, Raman scattering, and two-photon recombination 
rates.

Throughout this paper, we use a background cosmology with $\Omega_mh^2=0.13$, $\Omega_bh^2=0.022$, $T_{\rm CMB}=2.728\,$K, helium mass fraction 
$Y=0.24$, and an effective number of massless neutrinos $N_\nu=3.04$.  This implies the derived parameters $f_{\rm He}=0.0795$ (He:H ratio by number) 
and radiation density $\Omega_rh^2=4.196\times 10^{-5}$.  (These parameters are no longer up-to-date but should suffice to illustrate the physics and 
the magnitude of the corrections due to two-photon transitions.)
The cosmological constant is neglected because it has no influence on the recombination era, 
$\Omega_\Lambda(z)\le 10^{-8}$.  Reduced matrix elements are normalized following the convention of Berestetskii et~al. 
\cite{1980MINTF...4.....B}.
We will define a two-photon decay from a highly excited ($n\ge 3$) level to be ``\subla'' if both emitted photons are 
below the Ly$\alpha$ frequency and ``\supla'' if one photon is emitted above the Ly$\alpha$ frequency, in order to avoid repeating these cumbersome 
descriptions throughout the paper.

\section{The standard multi-level atom}
\label{ss:mla}

In this section, we revisit the basic equations of the multi-level atom (Section~\ref{ss:b1}) and the steady-state approximation (Section~\ref{ss:b2}).  

Most of the notation of this section is conventional.  We define $n_{\rm H}$ to be the total physical density of hydrogen nuclei (units of 
cm$^{-3}$) and define the abundance of hydrogen atoms in level $i$ relative to this total: $x_i=n_i/n_{\rm H}$.  A similar definition is used for 
free electrons and protons, $x_e=n_e/n_{\rm H}$ and $x_p=n_p/n_{\rm H}$.  We denote the energy of a bound state by $E_i<0$ and the energy of a free 
electron by $E_e>0$.
The degeneracies of states are denoted by $g_i$; in the case of hydrogen 
where we do not resolve fine structure levels, 
$g_{nl}=2(2l+1)$.  Einstein coefficients are denoted $A_{ji}$ for a transition from upper level $j$ to lower level $i$, and transition 
energies/frequencies are denoted $E_{ji}\equiv E_j-E_i$ and $\nu_{ji}=E_{ji}/h$.  The photon phase space density is $f(E)$.  When considering the phase 
space density 
just above (blueward of) or below (redward of) a line, we use the notation
\begin{equation}
f_{ji\pm} = f(E_{ji}\pm\epsilon),
\end{equation}
where $\epsilon$ is the line width, which is assumed to be infinitesimal (i.e. the time required to redshift through the line is assumed to be much 
less than the duration of recombination).  Temperatures are quoted in energy units, so that Boltzmann's constant is equal to 1.

\subsection{Basic equations}
\label{ss:b1}

The evolution equation for $x_i$ is
\begin{equation}
\dot x_i = \dot x_i|_{\rm bb} + \dot x_i|_{\rm bf} + \dot x_i|_{2\gamma},
\label{eq:xifull}
\end{equation}
where we have separated the net production of level $i$ into terms coming from radiative bound-bound, radiative bound-free, and two-photon 
processes.

The bound-bound term is
\begin{eqnarray}
\dot x_i|_{\rm bb} &=& \sum_{j>i} A_{ji}P_{ji}\left[(1+f_{ji+})x_j-\frac{g_j}{g_i}f_{ji+}x_i\right]
\nonumber \\ && + \sum_{j<i} 
A_{ij}P_{ij}\left[\frac{g_i}{g_j}f_{ij+}x_j-(1+f_{ij+})x_i\right],
\label{eq:xibb}
\end{eqnarray}
where the sums are over levels $j$ that are above ($j>i$) or below ($j<i$) the energy of level $i$; $A_{ji}$ is the Einstein coefficient; $g_i$ and 
$g_j$ are the level degeneracies; and $f_{ji+}$ is the photon phase space density on the blue (incoming) side of the line with frequency 
$\nu_{ji}=(E_j-E_i)/h$, which is necessary for following absorption and stimulated emission.  Here $P_{ji}$ is the Sobolev escape 
probability for the $ji$ line, which is the probability that a photon emitted in the line will escape via redshifting without being re-absorbed.  
It is given by
\begin{equation}
P_{ji} = \frac{1-e^{-\tau_{ji}}}{\tau_{ji}}
\label{eq:prob}
\end{equation}
with the optical depth
\begin{equation}
\tau_{ji} = \frac{c^3n_{\rm H}}{8\pi H\nu_{ji}^3}
A_{ji}\left(\frac{g_j}{g_i}x_i-x_j\right);
\end{equation}
see e.g. Ref.~\cite{2000ApJS..128..407S} for a derivation.
In practice the optical depth is $\tau_{ji}\ll1$ and $P_{ji}\approx 1$ except for the Lyman-series lines.  The jump condition for the radiation field 
across the line is obtained by finding the net rate of radiative de-excitations (i.e. photons emitted in the line), and multiplying by the conversion 
factor $8\pi H \nu_{ji}^3/(c^3n_{\rm H})$, which is the number of photon modes that redshift through the line per hydrogen nucleus per unit time:
\begin{equation}
f_{ji-} - f_{ji+} = \frac{8\pi H \nu_{ji}^3}{c^3n_{\rm H}}A_{ji}P_{ji}\left[x_j(1+f_{ji+}) - \frac{g_j}{g_i}x_if_{ji+}\right].
\label{eq:f-}
\end{equation}

The bound-free term includes spontaneous and stimulated recombination, and the inverse process, photoionization (see e.g. Sec. 2.3.1 of 
Ref.~\cite{2000ApJS..128..407S}).  It is
\begin{eqnarray}
\dot x_i|_{\rm bf} &=& \int \alpha_i(E_e) \Bigl\{P_{\rm M}(E_e) n_{\rm H}x_ex_p[1+f(E_e-E_i)]
\nonumber \\ &&
- \frac{dg}{dE_e\,dV} \frac{x_i}{g_i}f(E_e-E_i) \Bigr\} dE_e,
\end{eqnarray}
where $E_e$ is the energy of the free electron (more accurately, the center-of-mass energy of the electron and proton), $f(E_e-E_i)$ is the photon 
phase 
space density at specified energy, $\alpha_i(E_e)$ is the recombination 
coefficient to level $i$ for an electron and proton, and $P_{\rm M}(E_e)$ is the Maxwellian probability distribution for the electron
energy at matter temperature $T_{\rm m}$:
\begin{equation}
P_{\rm M}(E_e)\,dE_e = \frac 2{\sqrt\pi}\,\frac{E_e^{1/2}}{T_{\rm m}^{3/2}}e^{-E_e/T_{\rm m}}\,dE_e.
\end{equation}
The $1+f(E_e-E_i)$ factor accounts for stimulated recombination.
The factor $dg/(dE_e\,dV)$, i.e. the number of available electron states per unit volume per unit energy, is required in order to satisfy the principle 
of detailed balance.  It is equal to
\begin{equation}
\frac{dg}{dE_e\,dV} = \frac{8\sqrt 2\,\pi}{h^3}\mu^{3/2}E_e^{1/2},
\end{equation}
where $\mu$ is the proton-electron reduced mass, $\mu=m_e/(1+m_e/m_p)$.  Thus we may write:
\begin{eqnarray}
\dot x_i|_{\rm bf} &=& \frac 2{\sqrt\pi}\int \alpha_i(E_e) \frac{E_e^{1/2}}{T_{\rm m}^{3/2}} 
\nonumber \\ && \times \Bigl\{
n_{\rm H}x_ex_p[1+f(E_e-E_i)]e^{-E_e/T_{\rm m}}
\nonumber \\ && - \frac{2(2\pi T_{\rm m})^{3/2}}{h^3}\mu^{3/2}E_e^{1/2}\frac{x_i}{g_i} f(E_e-E_i)\Bigr\} dE_e.\;\;\;\;
\label{eq:xibf}
\end{eqnarray}
Finally the usual two-photon term is
\begin{equation}
\dot x_{2s}|_{2\gamma} = -\dot x_{1s}|_{2\gamma} = \Lambda_{2s}\left( -x_{2s} + x_{1s}e^{-E_{2s,1s}/T_{\rm r}}\right),
\label{eq:xi2g}
\end{equation}
where $\Lambda_{2s}$ is the two-photon decay rate from the $2s$ level, and the second term accounts for detailed balance.

Note that the bound-bound radiative transition rate requires knowledge of the photon phase space density $f(E_{ji+})$ on the blue side of the line.  
Seager et~al. \cite{2000ApJS..128..407S} assumed that this is simply a blackbody function,
\begin{equation}
f(E_{ji+}) = \frac 1{e^{E_{ji}/T_{\rm m}}-1}.
\end{equation}
In reality the phase space density for the Lyman-series lines will be greater than the blackbody value, because photons from the Ly$(n+1)$ line 
will redshift downward and add to the Ly$n$ line \cite{2007astro.ph..2143S,2007A&A...475..109C}.  In the absence of any processes that operate
between the Ly$n$ and Ly$(n+1)$ lines, one may simply use phase space conservation along a photon trajectory:
\begin{equation}
f_{np,1s+}(z) = f_{(n+1)p,1s-}\left[\frac{1-(n+1)^{-2}}{1-n^{-2}}(1+z)-1\right].
\label{eq:feedback}
\end{equation}
This equation refers to a photon phase space density at some earlier time.  We generate a lookup table of $f_{(n+1)p,1s-}$ as we proceed through 
recombination, and when a value is required we obtain it by 3-point quadratic interpolation.  This method requires that the step size $\Delta a/a$ be 
smaller than the spacing of the Lyman lines, $\nu_{(n+1)p,1s}/\nu_{np,1s}-1$, which imposes a significant but tractable computational burden.

The matter evolution equation is
\begin{equation}
\dot T_{\rm m} = -2HT_{\rm m} + \frac{8x_e\sigma_{\rm T}a_{\rm r}T_{\rm r}^4}{3(1+f_{\rm He}+x_e)m_ec}(T_{\rm r}-T_{\rm m}),
\end{equation}
where the second term comes from electron scattering \cite{2000ApJS..128..407S}.  It contains the Thomson cross section $\sigma_{\rm T}$ and the 
radiation constant $a_{\rm r}$.  The second 
term is very effective at driving the matter temperature toward the radiation temperature at early times.  Later when $x_e$ and $T_{\rm r}^4$ 
decline the matter temperature falls below the radiation temperature.  Since the radiation temperature scales as $T_{\rm r}\propto 
a^{-1}$, we may write this equation as
\begin{equation}
\frac{d}{dt}\left(\frac{T_{\rm m}}{T_{\rm r}}\right)
= -H\frac{T_{\rm m}}{T_{\rm r}} + \frac{8x_e\sigma_{\rm T}a_{\rm r}T_{\rm r}^4}{3(1+f_{\rm He}+x_e)m_ec}\left(1-\frac{T_{\rm m}}{T_{\rm r}}\right).
\label{eq:tmdot}
\end{equation}

\subsection{Steady-state approximation}
\label{ss:b2}

We now introduce the steady-state approximation, which is used to accelerate solution of the hydrogen recombination problem by eliminating the need for 
a stiff differential equation solver.

The above equations for the excited hydrogen levels ($i\neq 1s$) can be written in the form
\begin{equation}
\dot x_i = s_i + \sum_{j\neq 1s} (R_{ji}x_j-R_{ij}x_i) - \beta_ix_i - \gamma_ix_i,
\label{eq:xidot-1}
\end{equation}
where $s_i$ is a source term that depends on excitations from $1s\rightarrow i$ and recombinations to level $i$, and $R_{ij}$ is the transition rate 
from level $i$ to level $j$, $\beta_i$ is the rate of ionizations from level $i$, and $\gamma_i$ is the rate of decays 
(spontaneous, stimulated emission) from level $i$ to the ground state.  The coefficients $\{s_i,R_{ij},\beta_i,\gamma_i\}$ in general 
depend on the ionization fraction, radiation and matter temperatures, free electron density, and spectral distortions (since they involve $f_{ji+}$).  
The explicit expressions are: for the source term,
\begin{eqnarray}
s_i &=& \frac{g_i}2A_{i,1s}P_{i,1s}f_{i,1s+}x_{1s}
\nonumber \\ &&
      + \frac2{\sqrt\pi}n_{\rm H}x_ex_p\int [1+f(E_e-E_i)]
\nonumber \\ &&
      \times \frac{\alpha_i(E_e)E_e^{1/2}}{T_{\rm m}^{3/2}}e^{-E_e/T_{\rm m}}\,dE_e
\nonumber \\ &&
      + x_{1s}\delta_{i,2s}\Lambda_{2s}e^{-E_{2s,1s}/T_{\rm r}};
\label{eq:si}
\end{eqnarray}
for the transition term,
\begin{equation}
R_{ji} = 
       \left\{\begin{array}{lcl}
A_{ji}P_{ji}(1+f_{ji+}) & & j>i \\ A_{ij}P_{ij}f_{ij+}(g_i/g_j) & & j<i \end{array}\right.;
\label{eq:rji}
\end{equation}
for the ionization term,
\begin{equation}
\beta_i = \frac{2^{7/2}\pi\mu^{3/2}}{h^3} \int \alpha_i(E_e) \frac{E_e}{g_i} f(E-E_i) dE_e;
\end{equation}
and for the decay term,
\begin{equation}
\gamma_i = A_{i,1s}P_{i,1s}(1+f_{i,1s+}) + \delta_{i,2s}\Lambda_{2s}.
\label{eq:gammai}
\end{equation}

In these equations, there is a dependence on $x_e$, but almost no dependence on the $x_i$s.  The exceptions are (i) formally $x_{1s}=1-x_e-\sum_ix_i$ 
depends on the populations of the excited states, and (ii) the Sobolev probabilities $P_{ij}$ contain a small dependence on the excited states.  
However past analyses have shown that the populations of the excited states are very small (the maximum is $x_{2p}=5.5\times 10^{-14}$ at $z=1360$) so 
we may take $x_{1s}\approx 1-x_e$.  Also 
the Sobolev escape probabilities for transition between two excited state are always very close to 1 (the peak optical depth for such lines is 
$5.5\times 10^{-4}$ for 
H$\alpha$ at $z=1400$, corresponding 
to an escape probability of 0.99972).  Finally the Sobolev probability for transitions involving the $1s$ level are very small, but the high optical 
depth is due 
entirely to the $1s$ level -- the contribution from the excited level $np$ changes the optical depth by a factor of $x_{np}/3x_{1s}$ which has a 
maximum value of $1.7\times 10^{-12}$ (for $n=2$, $z=1600$; at later times or for higher $n$ it is lower).  This leads to a convenient re-packaging of 
Eq.~(\ref{eq:xidot-1}):
\begin{equation}
\dot x_i = s_i - \sum_j T_{ij}x_j,
\label{eq:xidot-2}
\end{equation}
where
\begin{equation}
T_{ij} = \delta_{ij}\left(\beta_i+\gamma_i+\sum_k R_{ik}\right) - R_{ji}
\end{equation}
is a square matrix.

The excited states of a hydrogen atom typically have very short lifetimes compared to the recombination timescale: the longest intrinsic lifetime 
is that of $2s$ ($\Lambda_{2s}^{-1}=0.12\,$s), and even taking into account Sobolev suppression the intrinsic lifetimes of the p-states are short, e.g. 
for $2p$ we have $(A_{2p,1s}P_{2p,1s})^{-1}\sim 1\,$s when the Ly$\alpha$ optical depth reaches its peak of $\sim 6\times 10^8$.  For comparison, 
hydrogen recombination takes several$\times 10^{12}\,$s.  Therefore we expect that to very high accuracy the excited level populations should be in 
steady state given Eq.~(\ref{eq:xidot-2}), i.e. we should have
\begin{equation}
x_i \approx \sum_j ({\bf T}^{-1})_{ij} s_j
\label{eq:xiss}
\end{equation}
and
\begin{equation}
\dot x_e \approx -\dot x_{1s} = -\sum_{i>1s} \left( \gamma_ix_i - \frac{g_i}2A_{i,1s}P_{i,1s}f_{i,1s+}x_{1s} \right).
\label{eq:xiss2}
\end{equation}
Formally it is the minimum eigenvalue $t_r^{-1}$ of the matrix ${\bf T}$ that indicates how rapidly a steady state solution is approached.  
During the redshift range covered by our code the relaxation time $t_r$ (reciprocal of the minimum eigenvalue of ${\bf T}$) has a maximum of $0.8\,$s, 
or $\sim 10^{-12}$ of the duration of recombination.  Thus the steady state approximation should very accurately describe the populations of the 
excited states and the 
ionization history as a function of time.  (It may not accurately describe the low-frequency spectral distortion since the latter depends on slight 
deviations of 
the excited state ratios from thermal equilibrium, i.e. one is taking differences of excited state populations that are nearly zero.  In general the 
spectral distortion is much more sensitive than the recombination history to numerical errors \cite{2006MNRAS.371.1939R, 2006A&A...458L..29C}.)

In the case of the Lyman lines where there is significant feedback, we need the phase space density $f_{(n+1)p,1s-}$, which is easily obtained from 
Eq.~(\ref{eq:f-}) once we have obtained $x_{(n+1)p}$ from Eq.~(\ref{eq:xiss}).

In order to eliminate the need for a stiff ODE solver we also need to remove the rapidly decaying mode from the matter temperature equation, 
Eq.~(\ref{eq:tmdot}).  At early times this can be done using the first-order asymptotic solution,
\begin{equation}
\frac{T_{\rm m}}{T_{\rm r}}(t)
= 1 - \frac{3(1+f_{\rm He}+x_e)m_ecH}{8x_e\sigma_{\rm T}a_{\rm r}T_{\rm r}^4}.
\label{eq:tmdot4}
\end{equation}
(See Appendix~\ref{app:matter} for a derivation).
This equation can be used to solve algebraically for $T_{\rm m}$ at each step in the integration, rather than requiring a stiff 
integrator.  In fact it is quite accurate at early times: the {\sc recfast} code, which fully integrates Eq.~(\ref{eq:tmdot}) at low redshifts, gives a 
value for $T_{\rm m}$ differing by $<0.01$\% from Eq.~(\ref{eq:tmdot4}) for all $z\ge 700$.  (At early times, {\sc recfast} sets $T_{\rm m}=T_{\rm 
r}$ since the Compton equilibrium time is fast, but we have checked that at $z\le 850$, {\sc recfast} has switched on the integration of the $\dot 
T_{\rm m}$ equation so this is a fair comparison.)
Accurate calculations at lower redshift would require one 
to ``switch on'' the differential equation for $T_{\rm m}$ once the matter-radiation equilibration time $(HJ)^{-1}$ becomes a significant fraction of 
the recombination time; at this point it would be possible to follow $T_{\rm m}$ with a non-stiff integrator.

In most implementations of the MLA, one first groups states together and then solves either the ``full'' Eq.~(\ref{eq:xifull}) or the steady state 
problem, Eq.~(\ref{eq:xiss2}).  For example the original calculation of Peebles \cite{1968ApJ...153....1P} used a single, lumped excited state and the 
steady state approximation (with some additional approximations such as neglect of stimulated recombination).

\section{Extension to two-photon transitions}
\label{sec:b3}

Now we would like to extend our analysis to include two-photon decays from the high-lying levels.  The selection rules for two-photon electric dipole 
decays restrict us to excited levels $ns$ and $nd$ (two-photon $np\rightarrow 1s$ decays are parity-forbidden).  At first sight this is a simple 
problem.  The two-photon decay rates can be calculated in tree-level QED, as done in Appendix~\ref{app:rates}, and the total transition rate 
$\Lambda_{nl}$ can be found using Eq.~(\ref{eq:lnl}).  This rate can be added to the evolution equations in analogy to Eq.~(\ref{eq:xi2g}).

Unfortunately this program does not work because the desired rates of e.g. spontaneous two-photon decay $3d\rightarrow 1s$ double-count the ``1+1'' or 
cascade decay process $3d\rightarrow 2p\rightarrow 1s$, and do not correct for re-absorption of the emitted photons.  We develop our understanding of 
this issue by first discussing the two-photon decay rates in Sec.~\ref{ss:2grates} and then reviewing previous attempts to incorporate two-photon 
transitions in the MLA (Sec.~\ref{ss:first}).  We discuss the physical resolution of the double-counting and re-absorption problems in 
Sec.~\ref{ss:physics}.  We then write down the full radiative transfer equations in Sec.~\ref{ss:eqn}.  The principal result of this paper will be the 
solution of this last set of equations.

\subsection{Rate coefficients}
\label{ss:2grates}

The theory and computation of two-photon decay rate coefficients is reviewed in Appendix~\ref{app:rates}.  Here we merely recount the most important 
facts and results.  The two-photon decay is
\begin{equation}
{\rm H}(nl) \rightarrow {\rm H}(1s) + h\nu + h\nu',
\end{equation}
where we define $\nu$ to be the higher-energy photon and $\nu'$ to be the lower-energy photon, so that there is no issue of indistinguishable particles 
and the corresponding factors of 1/(2!).  The two photons have frequencies that sum to $(1-n^{-2}){\cal R}$ due to conservation of energy.  The rate of 
decays is given by an integral of the differential rate
\begin{equation}
\Gamma_{nl\rightarrow 1s} = \int_{(1-n^{-2}){\cal R}/2}^{(1-n^{-2}){\cal R}} \frac{d\Gamma}{d\nu} d\nu,
\label{eq:integ-gamma}
\end{equation}
where in a thermal radiation bath the actual decay rate is related to the spontaneous rate $d\Lambda/d\nu$ by
\begin{equation}
\frac{d\Gamma}{d\nu} = \frac{d\Lambda}{d\nu} (1-e^{-h\nu/T_{\rm r}})^{-1} (1-e^{-h\nu'/T_{\rm r}})^{-1}.
\label{eq:dgdnu-b3}
\end{equation}
The increase over the vacuum ($T_{\rm r}=0$) rate is due to stimulated two-photon decays \cite{2006A&A...446...39C}.

\begin{figure*}
\includegraphics[angle=-90,width=6.5in]{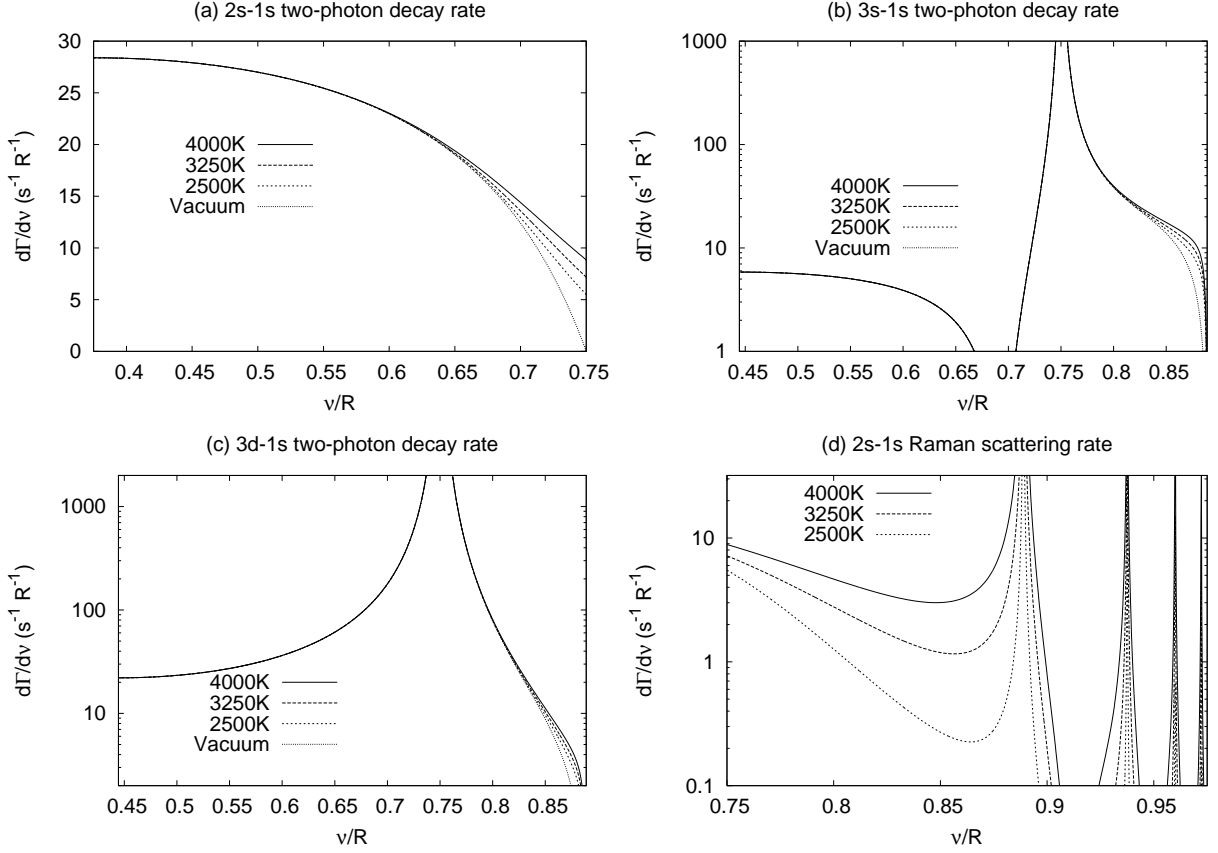}
\caption{\label{fig:rates}The two-photon transition rates for the most important excited levels of \HI.  The horizontal axis shows the frequency of 
the higher-energy photon (for $2\gamma$ decays) or the outgoing photon (for $2s\rightarrow 1s$ Raman scattering).  The differential transition rate 
$d\Gamma/d\nu$ is shown at several different temperatures and in vacuum
($T_{\rm r}=0$).  The two-photon spectra are shown from half the maximum frequency ($\nu_{\rm max}/2$) to $\nu_{\rm max}$ and are 
symmetric.  The Ly$\alpha$ resonance at $\nu=0.75{\cal R}$ is easily seen in the $3s,3d\rightarrow 1s$ rates.}
\end{figure*}

Panel (a) of Fig.~\ref{fig:rates} shows the two-photon decay rate from $2s\rightarrow 1s$ at several temperatures.  The differential decay 
rate $d\Gamma/d\nu$ is well-defined and integrable at all temperatures, and this decay presents no special problems.

Panel (b) of Fig.~\ref{fig:rates} shows the $3s\rightarrow 1s$ decays.  Here Eq.~(\ref{eq:dgdnu-b3}) contains a resonance at $\nu=3{\cal R}/4$, i.e. 
the Ly$\alpha$ frequency.  This corresponds to the pole in the matrix element from the $2p$ intermediate state.  Taking the two-photon decay formula at 
face value the total rate $\Gamma_{3s\rightarrow 1s}$ is infinite due to the $\sim 1/\Delta\nu^2$ divergence in the rate.  If one includes the 
imagniary energy (``pole displacement'') of $2p$ due to its finite lifetime ($1.6\,$ns) then one gets a finite decay rate, but very large: in vacuum 
$\Gamma_{3s\rightarrow 1s} = 6.3\times 10^6\,$s$^{-1}$.  This is equal to the one-photon transition rate $A_{3s,2p}$ (see discussion in CS08) for a 
simple physical reason: the sequential or ``1+1'' decay $3s\rightarrow 2p\rightarrow 1s$ {\em is} simply a two-photon decay with the photons on 
resonance.  (The one- and two-photon contributions can be separated as different diagrams that contribute to the width or imaginary energy shift of the 
$3s$ level $\Im \Delta E_{3s}$ \cite{2007JPhA...40..223J, 2008JPhA...41o5307J}, but $\Im \Delta E_{3s}$ does not give us the frequency distribution of 
emitted photons, which we need in our calculations.)  A similar problem occurs for the $3d$ initial level, as seen in panel (c) of 
Fig.~\ref{fig:rates}.

\begin{figure}
\includegraphics[angle=-90,width=3.2in]{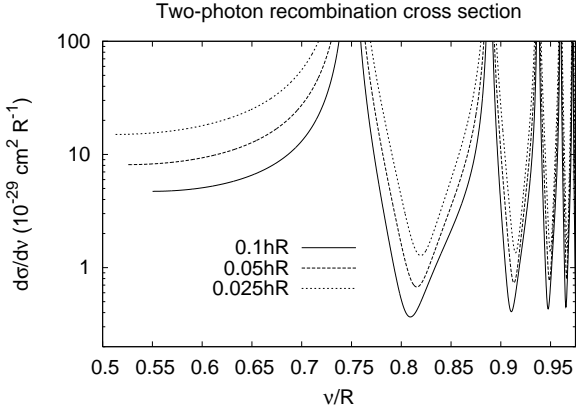}
\caption{\label{fig:2prec}
The differential cross section $d\sigma/d\nu$ for two-photon recombination at three incident electron energies (0.1, 0.05, and 0.025$h{\cal R}$).}
\end{figure}

A similar problem occurs for Raman scattering:
\begin{equation}
{\rm H}(nl) + h\nu' \rightarrow {\rm H}(1s) + h\nu.
\end{equation}
This time it is the difference of frequencies that is fixed by energy conservation: $\nu=\nu'+(1-n^{-2}){\cal R}$.
There is no Raman scattering in vacuum ($T_{\rm r}=0$) as it requires an initial-state photon, but one 
can define a Raman-scattering rate in analogy to Eq.~(\ref{eq:dgdnu-b3}) at finite temperature.  This is shown for the $2s$ initial state in the 
panel (d) of Fig.~\ref{fig:rates}.  In this case there are resonances when the virtual energy of the intermediate state corresponds with an 
energy level of hydrogen: at $\nu=8{\cal R}/9$, $15{\cal R}/16$, etc., and they reproduce the thermal rate for the sequential 1+1 photon transitions 
$2s\rightarrow 3p\rightarrow 1s$, $2s\rightarrow 4p\rightarrow 1s$, etc.

For completeness, we also discuss two-photon recombination,
\begin{equation}
{\rm H}^+ + e^- \rightarrow {\rm H}(1s) + h\nu + h\nu'.
\end{equation}
The differential cross sections $d\sigma/d\nu$ for this process are shown in Fig.~\ref{fig:2prec}.  It too exhibits 
Lyman-series resonances.  Note that the differential rate coefficient $\alpha^{(2)}(\nu,\nu')$ is equal to the incident electron velocity times 
$d\sigma/d\nu$: $\alpha^{(2)}(\nu,\nu') = v_e\,d\sigma/d\nu$.

At this stage, we have two basic problems.  One is the double-counting problem: we have already included 1+1 photon transitions in the MLA, so it seems 
incorrect to recount the resonant two-photon transitions.  The other is the re-absorption problem: the very large two-photon transition rates near 
resonance imply large optial depth for re-absorption of some of the emitted radiation, combined with the possibility that the Lyman-line spectral 
distortions could be immediately re-absorbed.  The next section critically discusses previous attempts to solve these problems.

\subsection{Past attempts at inclusion in MLA codes}
\label{ss:first}

The double-counting and re-absorption problems were recognized even in the very first attempts to include two-photon decays in the MLA.  There have 
been three major approaches suggested in the recombination literature so far: those of DG05, WS07, and CS08.  Here we describe each of them 
and explain why, while each of these methods contains important physical insights, none of them are both complete and applicable to the problem at 
hand.  In each of these cases the authors themselves noted that their solution was only a first approximation and that ultimately a 
radiative transfer calculation was required; in particular CS08 provided a detailed description of the problem.

DG05 avoided the double-counting problem by removing the 1+1 poles from the two-photon decay matrix element ${\cal M}$.  For example, in the case of 
the $3d\rightarrow 1s$ decay they removed the $2p$ intermediate state.  They then argued that in $ns,nd\rightarrow 1s$ decays, the largest contribution 
to the matrix element comes from the $np$ intermediate state.  Keeping only this term in ${\cal M}$ they arrived at a finite decay rate:
\begin{equation}
\Lambda^{\rm DG05}_{nl} \propto \frac{\nu_{nl,1s}^5}{2l+1}\left| \langle nl||{\bf r}||np\rangle\langle np||{\bf r}||1s\rangle \right|^2,
\end{equation}
which they normalize to the $2s$ recay rate $\Lambda_{2s}=8.22\,$s$^{-1}$.  They then substitute this rate coefficient into their MLA code in the same 
way that the traditional treatment inserts $\Lambda_{2s}$.  The DG05 rate coefficients scale as $\propto n$ and so a cutoff must be imposed; they 
impose such a cutoff when the wavelength of emitted radiation is less than the size of the atom ($\sim a_0n^2$) because at higher $n$ the electric 
dipole formula is no longer valid.

This approach obviously eliminates the double-counting problem.  Also by getting rid of the large differential rate $d\Lambda/d\nu$ near resonance it 
eliminates the re-absorption problem.  However it ignores the destructive interference in ${\cal M}$ that arises from consideration of the different 
intermediate states; it predicts $d\Lambda/d\nu\propto n$ for large $n$ and fixed $\nu$, whereas the actual scaling when interference is considered is 
$\propto n^{-3}$.  This occurs because while the largest contribution to the $ns,nd\rightarrow 1s$ matrix element comes from the $np$ intermediate 
state, the neighboring intermediate states $(n\pm 1)p$, $(n\pm 2)p$, etc. give opposite sign contributions and result in a cancellation of the matrix 
element that becomes more exact as $n\rightarrow\infty$ \cite{2007astro.ph..2144H}.  Thus to have even a qualitatively correct result for 
$d\Lambda/d\nu$ we must include the troublesome intermediate states such as $(n-1)p$, even if they are far from resonance.

WS07 presented an improved treatment of the problem.  They used rate coefficients for $3s\rightarrow 1s$ and $3d\rightarrow 1s$ two-photon decays that 
included all the $n\ge 3$ intermediate states (i.e. those that cannot be on-shell) \cite{1986PhRvA..33.1677C, 1988PhRvA..38.2189F}.  From the $n=3$ 
levels this gives a two-photon decay rate of 0.0664 times the DG05 estimate, a result of the $4p$, $5p$, etc. intermediate states partially cancelling 
the $3p$ contribution to ${\cal M}$.  WS07 did not have calculated rate coefficients for $n>3$ so they re-scaled all of the DG05 estimates by a factor 
of 0.0664 and included them in their MLA code.  The WS07 method is an improvement because it partially takes into account the destructive interference 
in the matrix element.  Nevertheless, it still does not take into account all intermediate states and retains the incorrect $d\Lambda/d\nu\propto n$ 
scaling.

CS08 presented the full calculation of the two-photon decay rate $d\Lambda/d\nu$ including all intermediate 
states.  However by including these intermediate states they re-introduce the double-counting and re-absorption problems.  CS08 argued that the 
double-counting problem should be solved by defining a ``1+1 decay profile'' ($\phi^{1+1\gamma}_{n_is/d}$ in their notation) consisting of the sum of 
Lorentzians at each resonance.  The decay rate from the 1+1 profile is $\int \phi^{1+1\gamma}_{n_is/d} d\nu = \sum_{n'=2}^{n-1}A_{nl,n'p}$.
They then subtract this from the full two-photon decay profile to give a ``pure'' two-photon profile.  Their effective 
two-photon rate is obtained by integrating this pure two-photon profile,
\begin{equation}
\Lambda^{\rm CS08}_{nl} = \int_{(1-n^{-2}){\cal R}/2}^{\nu_{\rm c}} \left( \frac{d\Lambda_{nl}}{d\nu} - \phi^{1+1\gamma}_{n_is/d} \right)d\nu,
\label{eq:cs07}
\end{equation}
where $\nu_{\rm c}$ is a cutoff frequency.  (CS08 denote this rate by $A^{2\gamma}_{n_is/d\rightarrow 1s}$.)  The ``obvious'' choice for the cutoff 
frequency would be the maximum frequency $(1-n^{-2}){\cal R}$, but CS08 argued that a lower cutoff should be used to solve the re-absorption problem.  
All photons emitted blueward of Ly$\alpha$ will be re-absorbed later in recombination.  Furthermore, photons emitted redward of Ly$\alpha$ will be 
re-absorbed if they are within the optically thick part of the red Ly$\alpha$ damping wing, so $\nu_{\rm c}$ should be somewhat below $\nu_{{\rm 
Ly}\alpha}$.  CS08 parameterized this in terms of the number of Doppler widths $x_{\Gamma,\rm c}$:
\begin{equation}
\nu_{\rm c} = \nu_{{\rm Ly}\alpha} + x_{\Gamma,\rm c} \sqrt{\frac{2T_{\rm r}}{m_{\rm H}c^2}}.
\end{equation}
(Note the sign convention that $x_{\Gamma,\rm c}<0$ corresponds to the red side of Ly$\alpha$.)  The pure two-photon rate $\Lambda^{\rm CS08}_{nl}$ 
depends only logarithmically on $x_{\Gamma,\rm c}$ because the 1+1 profile removes the order $(\nu-\nu_{{\rm Ly}\alpha})^{-2}$ term from the series
expansion of the integrand in Eq.~(\ref{eq:cs07}), so that at small $\nu-\nu_{{\rm Ly}\alpha}$ the leading term is of order $(\nu-\nu_{{\rm 
Ly}\alpha})^{-1}$, whose integral is logarithmically divergent.  CS08 considered several values of $x_{\Gamma,\rm c}$ ranging from $-10^4$ to $-10^5$, 
and found that their correction to the free electron abundance was $<0.5$\% in each case.  They also found that while the full 
two-photon profile $d\Lambda_{nl}/d\nu$ must be non-negative, there is no restriction on its sign after the 1+1 profile has been subtracted; their rate 
coefficient $\Lambda^{\rm CS08}_{nl}$ is positive for the $d$ initial states and negative for $s$ initial states.

While the CS08 treatment is the most complete thus far, there are three areas in which it could be improved.  One is that it contains a 
free parameter $x_{\Gamma,\rm c}$ that affects the results and it is not obvious which value should be used.  A second issue is the subtraction of the 
Lorentzian profile for the 1+1 decays.  While it is true that integrating the Lorentzian profile does give the one-photon formula 
$\sum_{n'=2}^{n-1}A_{nl,n'p}$ for the decay rate, the Lorentzian profile plays no role at all in the conventional MLA.  The derivation of the Sobolev 
approximation assumes that the line in question (here Ly$\alpha$) is a $\delta$-function, and hence it is valid for {\em any} line profile so long as 
it is sufficiently narrow.  Thus it is not clear mathematically why in Eq.~(\ref{eq:cs07}) we should use the Lorentzian as the ``already included'' 
profile instead of any other function that integrates to $\sum_{n'=2}^{n-1}A_{nl,n'p}$. A final distinct issue is that photons emitted blueward of 
Ly$\alpha$ (i.e. in \supla\ decays) will be re-absorbed, but {\em not instantaenously}, so they do have an opportunity to affect 
recombination; we will see later that they yield corrections of up to $\sim 1.5$\%.

\subsection{Physical resolution of the double-counting and re-absorption problems}
\label{ss:physics}

To resolve the double-counting and re-absorption problems, in the next section we will proceed to write down the two-photon radiative transfer 
equations.  A radiative transfer calculation is necessary to fully address re-absorption.  Both numerical solution of the two-photon equations 
(Sec.~\ref{sec:numeric}) and analytic approximations (Sec.~\ref{sec:semi}) will show that the two-photon decay rate and radiation spectrum both remain 
finite; the photon phase space density $f_\nu$ in the optically thick region near the Ly$\alpha$ line approaches a quasi-equilibrium value in which 
the 
$nl\rightarrow 1s$ decays and $1s\rightarrow nl$ absorptions roughly cancel.

The double-counting problem is conceptually trickier.  There is no physical distinction between a pure two-photon decay that happens to be on resonance 
and a 1+1 decay, so we are free to draw the distinction however we want so long as the total rate and the radiative transfer calculation are correct.  
Since the MLA is already treating the 1+1 decays and associated radiative transfer via the Sobolev approximation for the Lyman lines, it makes sense 
for us to set up the distinction in such a way as to preserve the validity of the Sobolev approximation.  The assumptions underlying the Sobolev 
approximation can be identified from e.g. the derivation in Refs.~\cite{1984mrt..book...21R, 2000ApJS..128..407S}; they are (i) that the line width is 
narrow enough that phase space factors, atomic density, and Hubble rate do not vary significantly during the passage of a photon through the line; (ii) 
the emission and absorption profiles are the same; and (iii) that the line is the only significant emission or absorption mechanism in its frequency 
range.  [Additionally the Ly$\alpha$ diffusion is neglected, but for very optically thick lines the photons in the line tend to come to 
quasi-equilibrium in which the phase space density is constant, $f\approx x_u/(g_ux_l/g_l-x_u)$.
In this case diffusion in frequency plays little role and the Sobolev escape probability works well.  This 
assumption will be revisited in future work where Ly$\alpha$ diffusion is included.]

Of these, (i) requires that the ``1+1'' decay profile to have support only within a frequency width $\Delta\nu\ll \nu_{{\rm Ly}\alpha}$.  Assumption 
(ii) is harder: the ratio of emission to absorption profile in a thermal medium is $\propto e^{-h\nu/T}$ if $h\nu\gg T$.  Of course the real Universe 
is not in thermal equilibrium, however the emission/absorption ratio exhibits this exponential dependence in the vicinity of Ly$\alpha$ so long as the 
excited levels $n\ge 2$ are in Boltzmann equilibrium with each other, even if the Saha equation is not obeyed and the ratio $x_{2p}/x_{1s}$ is not 
equal to the Boltzmann value.  [We will prove this explicitly in Sec.~\ref{ss:b} when we derive Eq.~(\ref{eq:f0n1}); the ratio of emission/absorption 
profiles {\em is} $f^0_\nu$.] In order for $e^{-h\nu/T}$ to vary by only a small fraction and satisfy assumption (ii), we need $\Delta\nu\ll T/h$.  
This ``$\ll$'' is especially stringent: the Sobolev escape probability is guaranteed to be valid to accuracy $\epsilon$ only so long as the 
emission/absorption ratio ($\propto e^{-h\nu/T}$) is constant to within $\epsilon$, so we need $\Delta\nu<\epsilon T/h$.  This 
condition may be relaxed if the two-photon absorption optical depth (dominated by $1s\rightarrow 3d$) redward of $\nu_1$ (the red boundary of the 
Ly$\alpha$ line) is $\gg 1$, since in this case all Ly$\alpha$ photons that escape the line are re-absorbed anyway and the escape probability is moot.  
This issue will be explored numerically when we vary $\nu_1$.

Assumption (iii) requires that 1+1 and pure two-photon decays not overlap in frequency -- that is, for each Lyman line, we must define a frequency 
range $\nu_1<\nu<\nu_2$.  If a two-photon decay produces a photon within this range, it gets counted as 1+1.  Otherwise (i.e. if the photon is far 
enough from resonance) it gets counted as a pure two-photon decay.  In this picture the ``1+1'' part of the profile resembles a Lorentzian truncated at 
the frequencies $\nu_1$ and $\nu_2$, and the far damping wings are considered pure two-photon decays, even if they resemble a Lorentzian shape.

As a practical matter there is another restriction on $\nu_1$ and $\nu_2$.  In our numerical code we are not including the pole displacement in the 
calculation of the two-photon transition rates, since this would introduce highly complicated temperature dependence into $d\Lambda_{nl}/d\nu$.  
Therefore we must choose the detunings $\nu_{{\rm Ly}\alpha}-\nu_1$ and $\nu_2-\nu_{{\rm Ly}\alpha}$ to be large compared to the intrinsic width 
$\Gamma_{2p}$ of the 
Ly$\alpha$ line.  This has the ancillary advantage that the 1+1 decay rate is not significantly reduced by the truncation of the damping wings: the 
fraction of decays that proceed through the far damping tails is
\begin{eqnarray}
f_{\rm far} &=& 1-\int_{\nu_1}^{\nu_2} \frac{\Gamma_{2p} d\nu}{4\pi^2[(\nu-\nu_{{\rm Ly}\alpha})^2 + (\Gamma_{2p}/4\pi)^2]}
\nonumber \\
&\approx& \frac{\Gamma_{2p}}{4\pi^2}\left( \frac 1{\nu_{{\rm Ly}\alpha}-\nu_1} + \frac 1{\nu_2-\nu_{{\rm Ly}\alpha}} \right)
\end{eqnarray}
in the Lorentzian approximation.  This quantity is $\ll 1$ so long as $\nu_{{\rm Ly}\alpha}-\nu_1$ and $\nu_2-\nu_{{\rm Ly}\alpha}$ are 
$\gg\Gamma_{2p}$.

In principle the Ly$\alpha$ line profile is affected at some level by two-photon corrections even within the range $\nu_1<\nu_{{\rm Ly}\alpha}<\nu_2$, 
because the decay process that emits the line (e.g. $3d\rightarrow 2p\rightarrow 1s$) interferes with other pathways (e.g. $3d\rightarrow 3p\rightarrow 
1s$) that disrupt the approximation that there is a single pole in the matrix element (which leads to a Lorentzian).  This is not a problem for us: as 
long as assumptions (i)--(iii) above hold, we may use the Sobolev approximation for the central part of the Ly$\alpha$ line, which does not depend on 
the line profile.

Note that our separation into 1+1 and pure two-photon decays differs from CS08.  Also because $\nu_1$ and $\nu_2$ are arbitrary, one cannot uniquely 
define a total two-photon decay coefficient $\Lambda_{nl}$ for the two-photon decays.  This is not a problem since the ``total'' $\Lambda_{nl}$ will 
not appear in our equations and the notion of this object plays no role in our solution.

Our prescription, like that of CS08, is not unique: $\nu_1$ and $\nu_2$ are arbitrary.  The solution we derive can only be correct (and useful) if, 
within the restrictions above, we get the same answer regardless of our choice of these numbers.  This will be verified numerically, and we will also 
find that the analytic approximations we derive have well-defined limits for small $\nu_{{\rm Ly}\alpha}-\nu_1$ and $\nu_2-\nu_{{\rm Ly}\alpha}$ that 
agree with the numerical results.

\subsection{Two-photon radiative transfer equations}
\label{ss:eqn}

In the last two sections we highlighted the reasons why it is difficult to graft two-photon transitions onto the MLA.  We will now write down the full 
set of two-photon equations, including those for the radiation field.  This full set of equations must include both two-photon emission and absorption 
in order to be consistent with the laws of thermodynamics, and should serve as a starting point for any numerical solution or analytic approximation.  
After writing down the equations we will consider two ways of implementing them in an MLA.

In the case of a two-photon decay from $nl\rightarrow 1s$, the net number of decays per unit photon frequency per unit time per unit H atom is
\begin{equation}
\Delta_{nl}(\nu) = \frac{d\Lambda_{nl}}{d\nu} \left[ (1+f_\nu)(1+f_{\nu'}) x_{nl} - \frac{g_{nl}}{g_{1s}} f_\nu f_{\nu'} x_{1s} \right],
\label{eq:dnl}
\end{equation}
so that the net two-photon decay rate is
\begin{equation}
\dot x_{nl}|_{2\gamma} = -\int_{(1-n^{-2}){\cal R}/2}^{(1-n^{-2}){\cal R}} \Delta_{nl}(\nu)\,d\nu
\label{eq:xnl}
\end{equation}
for $n\ge 2$, and
\begin{equation}
\dot x_{1s}|_{2\gamma}=-\sum_{nl,n\ge2}\dot x_{nl}|_{2\gamma}.
\end{equation}

In the case of the $2s\rightarrow 1s$ transition, these equations replace Eq.~(\ref{eq:xi2g}).  Note that it is only necessary to follow the 
higher-energy photon ($\nu>\nu'$): the lower-energy photon has $\nu'<{\cal R}/2$, and hence has negligible probability of re-exciting a hydrogen atom.  
We can calculate the probability of these photons interacting again by considering their optical depths for various processes. The optical depth per 
Hubble time for two-photon absorption is typically in the range $10^{-20}$--$10^{-16}$ at frequency $\nu_{{\rm Ly}\alpha}/2=3{\cal R}/8$.  The optical 
depth per Hubble time for Balmer continuum absorption is higher, reaching a peak of $2\times 10^{-4}$ at the threshold ${\cal R}/4$ and $z=1400$.  The 
optical depth in the H$\alpha$ line is $5.5\times 10^{-4}$ at $z=1400$, and since this is a resonance in the Raman scattering cross section the optical 
depth for Raman scattering is much less (typically by a factor of the resonance width divided by the detuning from resonance).  Since there are $\sim 
0.5$ net two-photon decays per hydrogen atom during H recombination, the latter numbers suggest that re-absorption of the lower-energy photon from the 
two-photon decays would influence recombination at the level of a few parts in $10^4$.  In practice the effect should be less since at $z\sim 1400$ the 
excited states H$(n\ge 2)$ and the continuum H$^++e^-$ are kept in equilibrium (in which case absorption or emission of Balmer photons has no effect) 
and at lower redshift the lower population of the $2s$ and $2p$ levels results in much smaller optical depths.  Therefore we have not followed the 
lower-energy ($\nu'$) photons.

Because some portions of the two-photon spectrum are optically 
thick, we will also need the equations for the photons.
For frequencies not on an \HI\ resonance line, the photons follow the equation
\begin{equation}
\dot f_\nu = H\nu\frac{\partial f_\nu}{\partial\nu} + \frac{c^3n_{\rm H}}{8\pi \nu^2}\sum_{nl} \Delta_{nl}(\nu),
\label{eq:dfnu}
\end{equation}
where the first term is the Hubble redshift, and the second term is the source (the coefficient describes the photon phase-space density).

One can write similar equations for two-photon recombination/ionization and Raman scattering.  Only those reactions involving the ground state are 
included because one can transition among the other states by optically thin one-photon transitions, which are much faster.
In the case of the Raman scattering reaction:
\begin{equation}
{\rm H}(nl) + h\nu' \rightarrow {\rm H}(1s) + h\nu,
\end{equation}
we extend Eq.~(\ref{eq:dnl}) by writing
\begin{equation}
\Delta_{nl}(\nu) = \frac{dK_{nl}}{d\nu} \left[ (1+f_\nu)f_{\nu'} x_{nl} - 
\frac{g_{nl}}{g_{1s}}f_\nu (1+f_{\nu'}) x_{1s} \right]
\end{equation}
for $\nu>(1-n^{-2}){\cal R}$ and $\nu'=\nu-(1-n^{-2}){\cal R}$.  Here $dK_{nl}/d\nu$ is the differential Raman scattering rate, defined by 
Eq.~(\ref{eq:k}).
Equation~(\ref{eq:xnl}) is extended to include Raman scattering by writing
\begin{equation}
\dot x_{nl}|_{\rm Raman} = \int_{(1-n^{-2}){\cal R}}^{\cal R} \Delta_{nl}(\nu)\,d\nu;
\end{equation}
we have only extended the integration up to ${\cal R}$ because if the output photon has a frequency above ${\cal R}$ then it immediately ionizes an H 
atom.
For two-photon recombination to the ground state, i.e.
\begin{equation}
{\rm H}^+ + e^- \rightarrow {\rm H}(1s) + h\nu + h\nu',
\end{equation}
one writes
\begin{eqnarray}
\Delta_c(\nu) &=& \int_0^\infty
\Bigl[ (1+f_\nu)(1+f_{\nu'}) n_{\rm H}x_ex_pP_{\rm M}(E_e)
\nonumber \\ &&
 - \frac{dg}{dE\,dV} f_\nu f_{\nu'} \frac{x_{1s}}2 \Bigr]
\alpha^{(2)}(\nu,\nu')\,dE_e,
\end{eqnarray}
where $h\nu'=E_e-E_{1s}-h\nu$, and $\alpha^{(2)}(\nu,\nu')$ is the differential two-photon recombination rate coefficient (in e.g. cm$^3$ s$^{-1}$ 
Hz$^{-1}$), given by Eq.~(\ref{eq:alpha2}).  The 
$\Delta_c(\nu)$ term is added along with the bound-bound two-photon transitions $\Delta_{nl}(\nu)$ in 
Eq.~(\ref{eq:dfnu}).

\section{Numerical approach}
\label{sec:numeric}

In Section~\ref{ss:eqn} we wrote down the radiative transfer and level population equations including two-photon transitions.  Now we solve them 
numerically by discretizing the two-photon continuum.  We first write down the methods of solution (Section~\ref{ss:nmethod}) and describe our fiducial 
choice of parameters (Section~\ref{ss:nparams}).  Then we consider the results obtained by this method in Section~\ref{ss:nresult}, turning on a 
sequence of effects in order: (A) the ``correct'' handling of the $2s\rightarrow 1s$ decay, including stimulated emission and feedback; (B) \subla\ 
two-photon decays, i.e. those from $n\ge 3$ where both photon frequencies are below Ly$\alpha$; (C) \supla\ two-photon decays, i.e. those where one 
photon frequency is above Ly$\alpha$; (D) Raman scattering; and (E) two-photon recombination/ionization.

\subsection{Method}
\label{ss:nmethod}

Our agenda here is to discretize the two-photon transfer equations.  We will do this by converting the continuous frequency distributions such as 
$d\Lambda_{nl}/d\nu$ into discrete distributions, i.e. lines.  We will define a set of virtual levels, such that a two-photon decay such as 
$3d\rightarrow 1s$ can be represented formally as a one-photon decay from $3d$ to a virtual level, and then from the virtual level to $1s$.  We will 
then write down a set of equations for these levels so that the equations for the occupation of the virtual level are formally equivalent to the 
two-photon radiative transfer equations.  We emphasize that this is {\em only} a formal correspondence used to force an MLA code to solve the 
two-photon transfer equations, and that the virtual levels are {\em not} actual levels of the hydrogen atom (they are not linearly independent states 
in the Hilbert space).
In order to be pedagogical, we will work through only the two-photon decay/excitation here, and then state the analogous results for Raman 
scattering and two-photon recombination/ionization.

In order for the discretization procedure to work, the total rate and (after appropriate smoothing) shape of the two-photon distribution must remain 
fixed.  Since two-photon absorption also plays a key role, we must also make sure that the two-photon optical depths and branching probabilities 
(relative probability of two-photon absorption to different excited levels) are preserved.

The simplest way of doing this is to multiply each probability distribution by a discretizing function $D'$:
\begin{equation}
\left.\frac{d\Lambda_{nl}}{d\nu}\right|_{\rm used} = \left.\frac{d\Lambda_{nl}}{d\nu}\right|_{\rm true}D'(\nu),
\end{equation}
with
\begin{equation}
D'(\nu) = \sum_{b=1}^{N_v} \Delta\nu_b\,\delta(\nu-\nu_b),
\end{equation}
with the ``spike'' frequencies ordered $\nu_1<\nu_2<...<\nu_{N_v}$ and with $\Delta\nu_b$ being the spacing between consecutive spikes.
One can also define the integral
\begin{equation}
D(\nu) = \int_{\nu_{\rm cut}}^\nu D'(\tilde\nu)\,d\tilde\nu \approx \nu-\nu_{\rm cut}.
\end{equation}
As always the delta function has infinitesimal width.
The frequency $\nu_{\rm cut}$ is 
the minimum frequency at which we follow the spectral distortion.
Note that $d\Lambda_{nl}/d\nu$ is only defined above half the maximum frequency, $\nu>(1-n^{-2}){\cal R}/2$; below this cutoff we will formally take 
$d\Lambda_{nl}/d\nu=0$.  We take $f_{\nu'}$ to be a blackbody here, i.e. we neglect the spectral distortion at $\nu'<(1-n^{-2}){\cal R}/2$.  As shown 
in Section \ref{ss:eqn}, this does not lead to any significant errors.

Now that we have a discretized spectrum, we proceed to solve the radiative transfer equations.  At frequencies in between spikes, Eq.~(\ref{eq:dfnu}) 
is simply a free streaming equation and has a solution analogous to Eq.~(\ref{eq:feedback}):
\begin{equation}
f(\nu_b+\epsilon,z) = f\left[\nu_{b+1}-\epsilon, \frac{\nu_{b+1}}{\nu_b}(1+z)-1\right].
\end{equation}
Within a spike the situation is more subtle.  At $\nu\approx\nu_b$, Eq.~(\ref{eq:dfnu}) becomes
\begin{eqnarray}
\dot f_\nu &=& H\nu_b\frac{\partial f_\nu}{\partial\nu} + \frac{c^3n_{\rm H}}{8\pi\nu_b^2} D'(\nu)
\nonumber \\ &&
  \times \Bigl\{\sum_{nl} (1+f_{\nu'})x_{nl}\frac{d\Lambda_{nl}}{d\nu}
\nonumber \\ &&
  - \sum_{nl} \left[\frac{g_{nl}}{g_{1s}}f_{\nu'}x_{1s}-(1+f_{\nu'})x_{nl}\right]\frac{d\Lambda_{nl}}{d\nu}f_\nu
  \Bigr\}.
\label{eq:inline}
\end{eqnarray}
Since the low-frequency photon comes from the CMB (i.e. we are neglecting two-photon absorption where the low-energy photon is part of the spectral 
distortion), we may take $f_{\nu'}$ to be constant across the spike (but of course it does depend on $n$).
Then we may define the coefficients
\begin{equation}
B_1 = \sum_{nl} \left[\frac{g_{nl}}{g_{1s}} f_{\nu'}x_{1s}-(1+f_{\nu'})x_{nl}\right]\frac{d\Lambda_{nl}}{d\nu}
\label{eq:b1}
\end{equation}
and
\begin{equation}
B_0 = \sum_{nl} (1+f_{\nu'})x_{nl}\frac{d\Lambda_{nl}}{d\nu}.
\end{equation}
Then Eq.~(\ref{eq:inline}) can be written as
\begin{equation}
\frac{\dot f_\nu}{H\nu_b} = -\frac{\partial f_\nu}{\partial\nu} + \frac{c^3n_{\rm H}}{8\pi H\nu_b^3}D'(\nu)(B_0-B_1f_\nu).
\end{equation}
This equation has an integral solution,
\begin{eqnarray}
f_\nu &=& f_{\nu_b+\epsilon} + \left(\frac{B_0}{B_1}-f_{\nu_b+\epsilon}\right)
\nonumber \\ && \times
\left( 1-\exp\left\{ \frac{c^3n_{\rm H}}{8\pi H\nu_b^3}B_1[D(\nu)-D(\nu_b+\epsilon)] \right\}\right)
\nonumber \\ &&
  + \int_\nu^{\nu_b+\epsilon} \frac{\dot f_{\tilde\nu}}{H\nu_b}
\nonumber \\ && \times
  \exp\left\{ \frac{c^3n_{\rm H}}{8\pi H\nu_b^3}B_1[D(\nu)-D(\tilde\nu)] \right\}
  d\tilde\nu.
\label{eq:inline2}
\end{eqnarray}
As we make the delta functions infinitesimally narrow, the integrand in the last term remains finite but the range of integrated variable becomes 
infinitesimal.  Therefore this term disappears.
We can simplify Eq.~(\ref{eq:inline2}) by defining
\begin{equation}
\tau_b(\nu) = -\frac{c^3n_{\rm H}}{8\pi H\nu_b^3}B_1[D(\nu)-D(\nu_b+\epsilon)],
\end{equation}
and
\begin{equation}
\Delta\tau_b\equiv \tau_b(\nu_b-\epsilon) = \frac{c^3n_{\rm H}}{8\pi H\nu_b^3}B_1\Delta\nu_b.
\end{equation}
Comparing with Eq.~(\ref{eq:b1}) this looks similar to a Sobolev optical depth; indeed, $\Delta\nu_b$ is the optical depth to two-photon absorption.

At the end of the day the relevant numbers to know are the net two-photon decay rate from each excited level via bin $b$, $\dot x_{nl}|_{2\gamma,b}$, 
and the photon phase space density at $\nu_b-\epsilon$.  The latter can be computed from Eq.~(\ref{eq:inline2}); it is given by
\begin{equation}
f_{\nu_b-\epsilon} = f_{\nu_b+\epsilon} + \left(\frac{B_0}{B_1}-f_{\nu_b+\epsilon}\right)(1-e^{-\Delta\tau_b}).
\label{eq:f-bin}
\end{equation}
The two-photon decay rate can be written as (see Eq.~\ref{eq:dnl})
\begin{eqnarray}
\dot x_{nl}|_{2\gamma,b}\!\! &=&\!\! -\int_{\nu_b-\epsilon}^{\nu_b+\epsilon}
\left[(1+f_\nu)(1+f_{\nu'})x_{nl} - \frac{g_{nl}}{g_{1s}}f_\nu f_{\nu'}x_{1s}\right]
\nonumber \\ && \!\!\times
\frac{d\Lambda_{nl}}{d\nu}D'(\nu)
 d\nu.
\end{eqnarray}
Using Eq.~(\ref{eq:inline2}) for the phase space density within the spike, and recalling that the last term is zero, we see that
\begin{equation}
\dot x_{nl}|_{2\gamma,b} = -\frac{d\Lambda_{nl}}{d\nu}
[(1+\bar f_{\nu_b})(1+f_{\nu'})x_{nl} - \bar f_{\nu_b} f_{\nu'}x_{1s}]
\Delta\nu_b,
\label{eq:dxbin}
\end{equation}
where the weighted mean phase space density has been defined as
\begin{equation}
\bar f_{\nu_b} = \frac{B_0}{B_1} + \left( f_{\nu_b+\epsilon}-\frac{B_0}{B_1}\right)\frac{1-e^{-\Delta\tau_b}}{\Delta\tau_b}.
\label{eq:fbar}
\end{equation}
It is convenient to define the Sobolev-like ``escape probability'' from the spike,
\begin{equation}
\Pi_b \equiv \frac{1-e^{-\Delta\tau_b}}{\Delta\tau_b}.
\end{equation}

In principle we could directly add Eqs.~(\ref{eq:f-bin}) and Eq.~(\ref{eq:dxbin}) to the level equations, since $\bar f_{\nu_b}$ is linear in the 
excited state populations so long as they are small enough that $x_{1s}$ dominates the sum in $B_1$ (Eq.~\ref{eq:b1}).  We want to do this in a way 
that is easily incorporated into the steady-state equations; this can be done by inventing a ``virtual'' level $b$ for each spike, and extending the 
steady-state source vector ${\bf s}$ and transition matrix ${\bf T}$ to include it in such a way as to reproduce the two-photon equations.  The 
populations of the virtual level for each spike will correspond directly to the spectral distortion, which will be convenient numerically.
Let us define for each spike an arbitrary ``degeneracy'' $g_b$.  Formally one should think of $g_b$ as infinitesimal, since the virtual levels cannot 
be occupied, but within the context of the steady-state approximation the argument below will work for any arbitrary value.
We will then define the virtual level occupation
\begin{equation}
x_b \equiv \frac{g_b}2 x_{1s} \bar f_{\nu_b}
\label{eq:xb}
\end{equation}
and the virtual level energy
\begin{equation}
E_b \equiv E_{1s}+h\nu.
\end{equation}
We will also define the virtual rate coefficient
\begin{equation}
A_{nl,b} = \left.\frac{d\Lambda_{nl}}{d\nu}\right|_{\nu=\nu_b}\Delta\nu_b.
\label{eq:a-virtual}
\end{equation}
Then Eq.~(\ref{eq:dxbin}) gives us the effect of a given spike on the rate equation for level $nl$.  Since $f_{\nu'}=f_{nl,b+}$ and 
$\bar f_{\nu_b}\ll 1$, this is
\begin{equation}
\dot x_{nl}|_b = -A_{nl,b}(1+f_{nl,b+})x_{nl}
  + \frac{g_{nl}}{g_b}A_{nl,b}f_{nl,b+} x_b.
\end{equation}
This is exactly equal to what Eq.~(\ref{eq:rji}) for $R_{ji}$ would give when plugged into the steady state equation, Eq.~(\ref{eq:xidot-1}).

In order to implement something like the steady state equations, we must also express the equation for $\bar f_{\nu_b}$ (Eq.~\ref{eq:fbar}) in language 
that allows us to define a ${\bf T}$-matrix.  That is, if we imagine extending Eq.~(\ref{eq:xidot-2}) to include both real ($r$) and virtual ($v$) 
levels,
\begin{equation}
{\bf 0} = \left(\begin{array}{c} {\bf s}^{r} \\ {\bf s}^v \end{array}\right)
- \left(\begin{array}{cc} {\bf T}^{rr} & {\bf T}^{rv} \\ {\bf T}^{vr} & {\bf T}^{vv} \end{array}\right)
\left(\begin{array}{c} {\bf x}^{r} \\ {\bf x}^v \end{array}\right),
\end{equation}
we want to construct values of ${\bf s}^v$, ${\bf T}^{vr}$, and ${\bf T}^{vv}$ that are equivalent to Eq.~(\ref{eq:fbar}).  (Note that we have already 
constructed the top row of the matrices.)  Re-writing Eq.~(\ref{eq:fbar}) in the form
\begin{eqnarray}
0 &=& -\left(\sum_{nl}f_{\nu'}\frac{g_{nl}}{g_b}A_{nl,b}\right)x_b
\nonumber \\ && +
 (1-\Pi_b) \sum_{nl} (1+f_{\nu'})A_{nl,b}x_{nl}
\nonumber \\ && +
\left(\sum_{nl}f_{\nu'}\frac{g_{nl}}{g_b}A_{nl,b}\right)\frac{g_b}{g_{1s}}x_{1s}f_{\nu_b+\epsilon}\Pi_b,
\end{eqnarray}
we can then define the virtual level source term
\begin{equation}
s_b = \left(\sum_{nl}f_{\nu'}\frac{g_{nl}}{g_b}A_{nl,b}\right)\frac{g_b}{g_{1s}}x_{1s}f_{\nu_b+\epsilon}
\frac{\Pi_b}{1-\Pi_b},
\end{equation}
the virtual-real sub-block of the ${\bf T}$-matrix,
\begin{equation}
T_{b,nl} = - \sum_{nl} (1+f_{\nu'})A_{nl,b},
\end{equation}
and the virtual-virtual sub-block,
\begin{equation}
T_{bb'} = \frac{\delta_{bb'}}{1-\Pi_b}\sum_{nl}f_{\nu'}\frac{g_{nl}}{g_b}A_{nl,b}.
\end{equation}
A simple algebraic calculation shows that this is precisely what one would calculate for $s_b$, $T_{b,nl}$, and $T_{bb'}$ if one were to 
make the replacement
\begin{equation}
A_{b,1s}P_{b,1s} \rightarrow \frac{\Pi_b}{1-\Pi_b}\sum_{nl} \frac{g_{nl}}{g_b}f_{\nu'}A_{nl,b}
\end{equation}
in Eqs.~(\ref{eq:si}) and (\ref{eq:gammai}).  This can also be re-written as
\begin{equation}
A_{b,1s}P_{b,1s} \rightarrow \frac{8\pi H\nu_b^3}{c^3n_{\rm H}} \frac{1-e^{-\Delta\tau_b}}{1-(1-e^{-\Delta\tau_b})/\Delta\tau_b}.
\label{eq:ap}
\end{equation}
In other words, the steady-state equations can be extended to include the two-photon decays by adding a 
set of virtual levels, inserting a virtual one-photon transition between each level $nl$ and a virtual level with Einstein coefficient given by the 
two-photon decay rate, and introducing a transition from the virtual level to the ground state $1s$ with rate given by Eq.~(\ref{eq:ap}).

The above methodology can be extended to include Raman scattering and two-photon recombination/ionization with no conceptual changes but with the 
addition of some new reaction rates.  The Raman scattering reaction, Eq.~(\ref{eq:raman}), is included by including a new radiative rate
\begin{equation}
A_{b,nl} = \frac{g_{nl}}{g_b} \frac{dK_{nl}}{d\nu}\Delta\nu_b,
\end{equation}
and two-photon recombination is included by writing an effective recombination coefficient to the virtual level:
\begin{equation}
\alpha_b(E_e) = \alpha^{(2)}(\nu,\nu')\Delta\nu_b,
\end{equation}
where $h\nu'=E_e-E_{1s}-h\nu$.
Equation~(\ref{eq:ap}) remains valid as long as we define a total optical depth
\begin{eqnarray}
\!\!\!\!\Delta\tau_b &=& \frac{c^3n_{\rm H}}{8\pi H\nu_b^3} x_{1s}\Bigl[
\sum_{nl>b}\frac{g_{nl}}{g_{1s}}f_{\nu'}\frac{d\Lambda_{nl}}{d\nu}
\nonumber \\ &&
+ \sum_{nl<b}\frac{g_{nl}}{g_{1s}}(1+f_{\nu'})\frac{dK_{nl}}{d\nu}
\nonumber \\ &&
+ \frac{2^{7/2}\pi\mu^{3/2}}{h^3} \int_0^\infty dE_e \alpha^{(2)}(\nu,\nu')\frac{E_e}{g_{1s}}f_{\nu'}\,dE_e
\Bigr].
\end{eqnarray}

\subsection{Parameters}
\label{ss:nparams}

We have run our numerical radiative transfer code several times with different sets of parameters.  These choices are summarized in 
Table~\ref{tab:params}.  The truncation frequencies of the Ly$\alpha$ line (outside of which two-photon decays are treated using the numerical 
radiative transfer instead of being treated as 1+1 decays with the Sobolev approximation) are $\nu_{1,2}=\nu_{{\rm Ly}\alpha}\pm\Delta\nu_{\rm max}$.  
In the latter case one should compare to the intrinsic width of the $2p$ intermediate state, $\Gamma_{2p}=A_{2p,1s}=6.3\times 10^8\,$s$^{-1}$ in 
vacuum.  (The additional finite-temperature broadening is due mainly to $2p\rightarrow 3s,3d$ transitions and for recombinaton-era temperatures is $\ll 
A_{2p,1s}$.)  Thermal equilibrium conditions were assumed for the initial data at $z_{\rm init}$.  This redshift was chosen to be after the completion 
of \HeII$\rightarrow$\HeI\ recombination in order to avoid needing to simultaneously follow hydrogen and helium.  The ``Basic'' set of parameters will 
be used for all plots in the paper unless otherwise indicated.  All runs commenced at $z_{\rm init}=1605.8$ and used $n_{\rm max}=30$ shells of the 
hydrogen atom (465 sublevels including $l$ resolution).

\begin{table}
\caption{\label{tab:params}The parameters for each run of the radiative transfer code.  Shown are the number of 
virtual levels $N_{\rm virt}$; and the step size in log scale factor $\Delta\ln a$; the maximum detuning from Ly$\alpha$ at which a photon is 
considered resonant, $\Delta\nu_{\rm max}$; the step size in the bound-free rate integrals, $\Delta\ln E_e$; and the line corresponding to the maximum 
frequency used in the grid.}
\begin{tabular}{lccccclcccccc}
\hline\hline
Run & & $N_{\rm virt}$ & & $10^5\Delta\ln a$ & & $\Delta\nu_{\rm max}$ & & $\Delta\ln E_e$ & & Max. grid \\
  & &  & &  & &  GHz & & & & frequency \\
\hline
Basic & & 367 & & 4.25\textcolor{white}0 & & \textcolor{white}0105 & & 0.1 & & Ly$\iota$ \\
WideLine1 & & 365 & & 4.25\textcolor{white}0 & & \textcolor{white}0315 & & 0.1 & & Ly$\iota$ \\
WideLine2 & & 365 & & 4.25\textcolor{white}0 & & \textcolor{white}0945 & & 0.1 & & Ly$\iota$ \\
LoRes & & 262 & & 4.25\textcolor{white}0 & & 6116\footnotemark[1] & & 0.1 & & Ly$\iota$ \\
HalfStep & & 367 & & 2.125 & & \textcolor{white}0105 & & 0.1 & & Ly$\iota$ \\
HiRes & & 695 & & 4.25\textcolor{white}0 & & \textcolor{white}0105 & & 0.1 & & Ly$\mu$ \\
CoarseBF & & 367 & & 4.25\textcolor{white}0 & & \textcolor{white}0105 & & 0.2 & & Ly$\iota$ \\
\hline\hline
\end{tabular}
\footnotetext[1]{For the red boundary $\nu_1$; the blue boundary is at $\nu_{{\rm Ly}\alpha}+10259\,$GHz.}
\end{table}

\comment{
For the Basic run, the frequency grid for the virtual levels was constructed as follows.  The grid must start at $3{\cal R}/8$ in order to follow the 
entire two-photon continuum from $2s$ (the higher-frequency photon emerges between $3{\cal R}/8$ and $3{\cal R}/4$).  We thus place the first 20 
virtual levels centers at 0.380859375(0.01171875)0.603515625 [the notation $x(y)z$ will represent equal steps from $x$ to $z$ with step size $y$], 
thereby uniformly covering the range from $3{\cal R}/8$ to $(1-1.6^{-2}){\cal R}$.  Beyond this
we write each grid frequency as
\begin{equation}
\nu_b = {\cal R}\left( 1 - \frac 1{n_b^2}\right),
\end{equation}
where $n_b$ is an effective fractional quantum number for the virtual level.  We must start at $n_b=1.6$ and the highest resolution is required near 
the Ly$\alpha$ line ($n=2$).  The next set of 20 virtual levels are at $n_b=1.605(0.01)1.795$; then we place 68 at $n_b=1.80125(0.0025)1.96875$; 60 at 
$n_b=1.9705(0.001)2.0295$; 68 at $n_b=2.03125(0.0025)2.19875$; 72 at $n_b=2.2125(0.025)3.9875$; and 30 at $n_b=4.1(0.2)9.9$.  The real hydrogen lines 
above $n_b=10$ (Ly$\iota=0.99{\cal R}$) are so closely spaced that in our Basic run we do not attempt to increase the frequency resolution in this 
region.
}

For the Basic run, the frequency grid extends only up to Ly$\iota=0.99{\cal R}$ since above this frequency the real hydrogen lines are very closely 
spaced and feedback is almost instantaneous.
The ``WideLine1/2'' runs are the same as the Basic run except that we increased $\Delta\nu_{\rm max}$ by a factor of 3 and 9 respectively, i.e. a 
two-photon decay had to emit a photon within 315 (instead of 105) GHz of the Ly$\alpha$ line in order to be considered a 1+1 decay instead of being 
treated with the two-photon radiative transfer.  Its purpose is to show that $\Delta\nu_{\rm max}$ does not matter.  The ``HalfStep'' run is a 
convergence test that used half the time step of the Basic run.  The ``LoRes'' run has a lower-resolution grid of virtual levels, while ``HiRes'' has a 
higher-resolution grid (twice the resolution except within 25$\,$THz of Ly$\alpha$) that extends all the way up to Ly$\mu$ ($1s-13p$).  The 
``CoarseBF'' run uses a 
step size of $\Delta\ln E_e=0.2$ instead of $0.1$ for evaluating the energy integrals for the bound-free rates.
All of the alternative runs gave changes in the ionization fraction $|\Delta x_e|/x_e<10^{-4}$, except WideLine2 and LoRes for which the maximum change
was $4\times 10^{-4}$.

\subsection{Results}
\label{ss:nresult}

\begin{figure}
\includegraphics[angle=-90,width=3.4in]{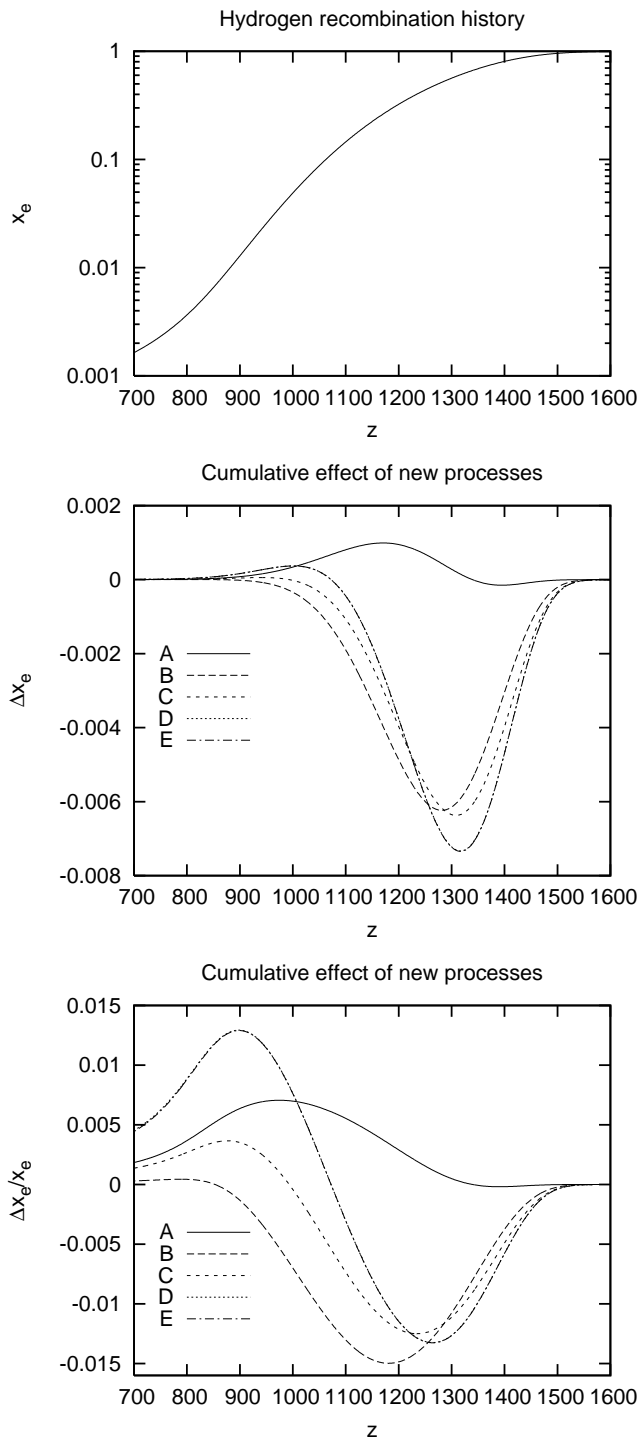}
\caption{\label{fig:mods}{\em Upper panel:} the recombination history. {\em Middle panel:} The absolute change $\Delta x_e$ caused by incorporation of 
each process. {\em Lower panel:}
The relative change $\Delta x_e/x_e$.  The changes are 
cumulative in the sense that the ``B'' curve represents the effect of incorporating A and B, etc.  The (D) and (E) curves are almost indistinguishable 
on the plot.}
\end{figure}

\begin{figure*}
\includegraphics[angle=-90,width=7.0in]{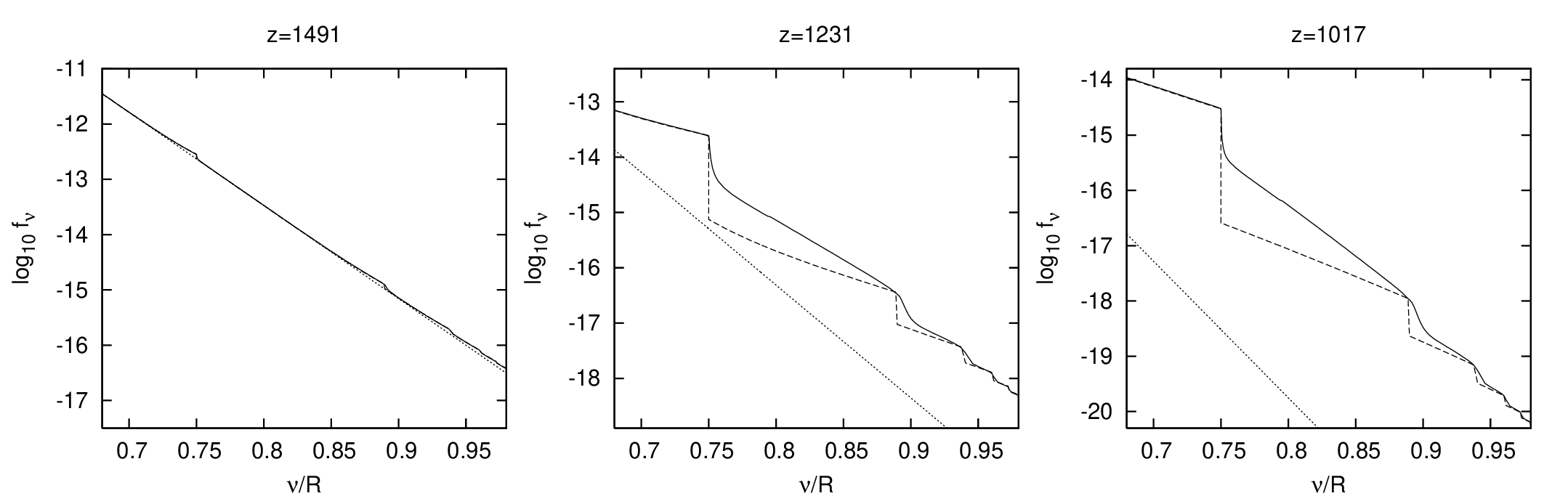}
\caption{\label{fig:phot}The radiation spectrum as a function of time.  Each panel shows the full spectrum (solid line); the spectrum neglecting the 
highly excited two-photon and Raman transitions (i.e. including only modification A) (dashed line); and the blackbody spectrum (dotted line).
We do not show the lower frequencies because our code does not follow the lower-energy photon in two-photon decays nor the Balmer, Paschen, etc. 
lines or bound-free continuum.}
\end{figure*}

The change in the recombination history caused by the new physics is shown in Figure~\ref{fig:mods}.  We will discuss the effects qualitatively here; 
analytic estimates will be given in Section~\ref{sec:semi} and compared with the numerical results.  The strength of the numerical result is that the 
only approximations made are those intrinsic to the two-photon radiative transfer equation and the
step sizes, frequency resolution, etc., which can be varied to establish convergence.  The analytic approximations yield 
insight but require numerous additional simplifying assumptions whose validity is difficult to assess.

The first effect that we have considered is the inclusion of stimulated two-photon decay and nonthermal two-photon absorption in the $2s\rightarrow 1s$ 
transition (A).  These effects go in opposite directions since stimulated emission increases the effective $2s\rightarrow 1s$ decay rate and hence 
accelerates recombination \cite{2006A&A...446...39C} whereas nonthermal two-photon absorption retards it \cite{2006AstL...32..795K}.  Initially the net 
effect is a slight acceleration because of the strong stimulated emission at high temperature, but by $z=1345$ this turns into a retardation ($\Delta 
x_e>0$) as enough hydrogen recombines to build a significant Ly$\alpha$ spectral distortion.  This grows to $\Delta x_e\approx 0.001$ at $z\approx 
1200$ and then tapers off.  For most redshifts we are in agreement with Ref.~\cite{2006AstL...32..795K} (compare their Figure 9 to our 
Figure~\ref{fig:mods}), but at the tail end of recombination ($z<800$) we find a smaller effect, $\Delta x_e/x_e=0.0018$ at $z=700$ instead of 0.0047.  
This is probably due to the approximation in Ref.~\cite{2006AstL...32..795K} that the excited levels $n\ge 2$ are in Saha equilibrium; if we force Saha 
equilibrium in our code, our $\Delta x_e/x_e$ rises to 0.0048.

Next we considered (B), the \subla\ two-photon transitions $ns,nd\leftrightarrow 1s$.
In these cases the emitted photons can never be resonantly absorbed by a hydrogen atom.  As one would intuitively expect, the addition of 
this new process brings the Universe closer to thermal equilibrium, that is it accelerates recombination. These transitions have a huge effect: they 
reduce the electron abundance by as much as $\sim 2$\%, and moreover this reduction is occurring at $z\sim 1100$, i.e. at the surface of last 
scattering where it matters most.

The next effect is (C), the \supla\ two-photon decays.  This is another process connecting excited states 
of \HI\ to the ground state and naively we would expect that like (B) it would accelerate recombination.  This is true at first.  However at $z=1290$ 
the situation reverses and the new process delays recombination.  The reason for this is that these two-photon decays are populating the region of 
photon phase space at $\nu>\nu_{{\rm Ly}\alpha}$ to far above the blackbody value.  In the later stages of recombination, these photons redshift down 
to Ly$\alpha$ and begin re-exciting hydrogen atoms.  This delayed feedback is present even in the standard recombination calculation because some 
photons escape from the Ly$\beta$ line \cite{2007A&A...475..109C} but the two-photon decays to frequencies between Ly$\alpha$ and Ly$\beta$ make a 
significant addition because their probability of immediate re-absorption is much lower.  A similar result occurs for (D) Raman scattering, which 
initially accelerates recombination but then slows it down.  In this case the main effect is $2s\rightarrow 1s$ scattering in which the incoming photon 
is at $\nu<\nu_{{\rm H}\alpha}$ and the outgoing photon is at $\nu_{{\rm Ly}\alpha}<\nu<\nu_{{\rm Ly}\beta}$.

Two-photon recombinations (E) have an accelerating effect as one would intuitively expect.  However since the electron occupation probability of the 
continuum states ($E_e>0$) is much less than the $2s$, $3s$, and $3d$ states this is a small effect; we find a maximum of $|\Delta x_e|/x_e\approx 
10^{-3}$.

We can also examine the effect on the high-frequency end of the spectral distortion.  In Fig.~\ref{fig:phot}, we show the photon spectrum at three 
epochs during recombination.  
The full calculation inlcuding effects A--E is shown compared to the case with effect A (modifications to $2s\leftrightarrow 1s$) only.  (We show the 
case with the $2s\leftrightarrow 1s$ modifications turned on because in the original case with the $2s\rightarrow 1s$ decays followed with the 
integrated rate coefficient $\Lambda_{2s}$ only, the code cannot follow the two-photon spectral distortion.)  The large jumps at each Lyman line are 
due to photons redshifting out of the line, as described by Peebles \cite{1968ApJ...153....1P}.  However even at early times there is a spectral 
distortion {\em between} the Lyman lines that is due to two-photon and Raman transitions.  Particularly impressive is the phase space between 
Ly$\alpha$ and Ly$\beta$: here the photon occupation number can be up to an order of magnitude larger than in the traditional treatment.  Due to their 
very long mean free paths, these photons can survive to $z<1100$ and slow down recombination by re-exciting hydrogen atoms.

\section{Analytic approach}
\label{sec:semi}

In the previous section we numerically evaluated the effect of two-photon transitions on hydrogen recombination using a radiative transfer scheme based 
on ``virtual'' levels.  Here we consider simple analytic estimates of their effect.  The main purposes are to provide an independent check of the 
fully numerical result and to develop some understanding of the magnitude of the various effects.  We consider the improvements to the 
$2s\leftrightarrow 1s$ decays (Section~\ref{ss:a}); \subla\ two-photon decays (Section~\ref{ss:b}); \supla\ two-photon decays 
(Section~\ref{ss:c}); and Raman scattering (Section~\ref{ss:d}).  We also considered the two-photon recombinations 
in the numerical code, however since they were found to be unimportant we will not construct an analytic estimate.

Our analytic investigation of the two-photon processes is based on many simplifying assumptions, most notably that (1) $T_{\rm r}\ll h{\cal R}$; (2) 
the low-lying excited states of the hydrogen atom ($n=2,3$) are in Boltzmann equilibrium; and (3) that the radiation field can be treated as being in 
steady state (or some lowest-order non-steady state approximation, in the cases of two-photon decays where one photon is above Ly$\alpha$ and Raman 
scattering).  These assumptions are only marginally satisfied during recombination and it is difficult to directly estimate the residual error. None of 
these approximations is made in the numerical code.  For these reason we believe the numerical results in the previous section are the most reliable 
estimate of the effect of two-photon transitions.

\subsection{Stimulated emission and nonthermal absorption in $2s\leftrightarrow 1s$}
\label{ss:a}

The $2s\leftrightarrow 1s$ transition is modified by both stimulated emission and absorption of the Lyman-$\alpha$ spectral distortion.  The approach 
described here, like that of Refs.~\cite{2006A&A...446...39C,2006AstL...32..795K}, assumes the optically thin limit for the $1s\rightarrow 2s$ 
absorption, but also takes the leading-order term in the power series for the spontaneous decay rate in order to arrive at an analytic formula.
This effect 
requires a thermal CMB photon to be present at one of the photon frequencies $\nu'$ and hence occurs in the regime $h\nu'\sim T_{\rm r}\ll h{\cal R}$.  
In this regime the $2p$ intermediate state dominates; using the exact hydrogenic matrix elements $\langle 1s||{\bf r}||2p\rangle = 2^{15/2}a_0/3^{9/2}$ 
and $\langle 2s||{\bf r}||2p\rangle = -3^{3/2}a_0$, we find (see Eq.~\ref{eq:lambda})
\begin{equation}
\frac{d\Lambda_{2s}}{d\nu} \approx \frac{512\alpha_{\rm fs}^6}{729} \frac{\nu'}{\cal R}.
\end{equation}
The total rate $\Delta \dot x_\downarrow^{\rm(A)}$ of stimulated emission minus absorption of the spectral distortion is
\begin{equation}
\Delta\dot x^{\rm(A)}_\downarrow = \int_0^{\nu_{{\rm Ly}\alpha}/2} \frac{d\Lambda}{d\nu} f_{\nu'}
(x_{2s} - x_{1s}\Delta f_\nu) d\nu',
\end{equation}
where $\Delta f_\nu$ is the phase space density of spectral distortion photons (i.e. the total phase space density with the blackbody subtracted).  
Since $f_{\nu'}$ is a blackbody function, we may write
\begin{equation}
\Delta\dot x^{\rm(A)}_\downarrow \approx \frac{512\alpha_{\rm fs}^6}{729{\cal R}}
\int_0^{\nu_{{\rm Ly}\alpha}/2}
\frac{\nu'(x_{2s} - x_{1s}\Delta f_\nu)
}{e^{h\nu'/T_{\rm r}}-1} 
d\nu'.
\end{equation}
Here $x_{2s}$ and $x_{1s}$ are independent of frequency, and $\Delta f_\nu$ varies smoothly with $\nu'$ (since we are looking on the red side of 
Ly$\alpha$; if we considered negative $\nu'$ there would be a jump).  Therefore this integral can be approximated using one-point Gaussian integration
(see e.g. Chapter 19 of Ref.~\cite{1986nmse.book.....H}.)
The idea of one-point Gaussian integration is that if one has an analytically integrable weight function $w(\nu)$, and a linear function 
$g(\nu')=a+b\nu'$, that:
\begin{equation}
\int w(\nu')g(\nu')\,dx = C_0g(\nu'_0),
\label{eq:gauss}
\end{equation}
where $C_0 = \int w(\nu')d\nu'$ and $\nu'_0 = C_0^{-1}\int \nu' w(\nu')d\nu'$.  This enables integration with a single sample of the function $g$ 
instead of two as is usual for linear functions.
For functions $g(\nu')$ that vary smoothly and are being integrated 
over a 
narrow range we may use Eq.~(\ref{eq:gauss}) as an approximation.  In our case,
$w(\nu') = \nu'/(e^{h\nu'/T_{\rm r}}-1)$ and $g(\nu') = x_{2s} - x_{1s}\Delta f_\nu$.  Application of Gaussian integration gives:
\begin{equation}
\int_0^{\nu_{{\rm Ly}\alpha}/2}
\frac{\nu' (x_{2s} - x_{1s}\Delta f_\nu) d\nu'
}{e^{h\nu'/T_{\rm r}}-1} 
\approx C_0(x_{2s} - x_{1s}\Delta f_{\nu_0}),
\end{equation}
where
\begin{equation}
C_0 = \int_0^\infty \frac{\nu' d\nu'}{e^{h\nu'/T_{\rm r}}-1} = \frac{\pi^2T_{\rm r}^2}{6h^2}
\end{equation}
and
\begin{equation}
\nu'_0 = C_0^{-1}\int_0^\infty \frac{\nu'{^2} d\nu'}{e^{h\nu'/T_{\rm r}}-1} = \frac{12\zeta(3)T_{\rm r}}{\pi^2h}.
\end{equation}
(The upper cutoff in the integral does not matter because in practice the exponential provides a cutoff.)  Here $\zeta(3)$ is the Riemann 
$\zeta$-function, and $\nu_0\equiv \nu_{{\rm Ly}\alpha}-\nu'_0$.  This leads to
\begin{equation}
\Delta\dot x^{\rm(A)}_\downarrow \approx \frac{256\pi^2\alpha_{\rm fs}^6T_{\rm r}^2}{2187h^2{\cal R}}(x_{2s} - x_{1s}\Delta f_{\nu_0}).
\label{eq:analytic-a}
\end{equation}

\begin{figure}
\includegraphics[angle=-90,width=3.2in]{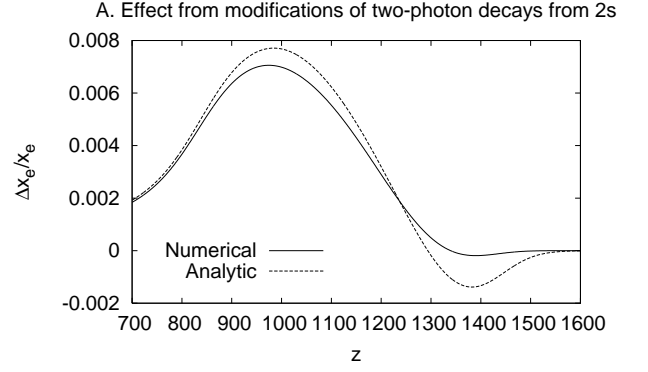}
\caption{\label{fig:diffa}A comparison of methods for handling the $2s\leftrightarrow 1s$ two-photon transitions.
We show the numerical radiative transfer method (solid line) and the analytic approximation of Eq.~(\ref{eq:analytic-a}) (dashed line).}
\end{figure}

We have compared Eq.~(\ref{eq:analytic-a}) to the full numerical result for the $2s\leftrightarrow 1s$ two-photon transitions in Fig.~\ref{fig:diffa}.

\subsection{\Subla\ two-photon decays}
\label{ss:b}

We now consider the \subla\ two-photon decays from higher excited states of \HI\ such as $3s$ and $3d$.
In most cases the higher-frequency photon is just below Ly$\alpha$ and can be cosidered to be part of the red damping tail.  Therefore in 
order to construct an analytic theory of the two-photon transitions we must develop an approximation for the radiative transfer in the red damping 
tail.  The approximation described here draws heavily on that of Appendix B of Hirata \& Switzer \cite{2007astro.ph..2144H}.

The two-photon transfer equation, Eq.~(\ref{eq:dfnu}) with Eq.~(\ref{eq:dnl}) used for $\Delta_{nl}(\nu)$, can be written as a linear equation for $f$:
\begin{equation}
\frac{\dot f_\nu}{H\nu} = \frac{\partial f_\nu}{\partial\nu} - \kappa_\nu (f_\nu - f^0_\nu),
\label{eq:2b1}
\end{equation}
where we have defined the functions
\begin{equation}
\kappa_\nu = \frac{c^3n_{\rm H}}{8\pi H\nu^3} \sum_{nl} \left[\frac{g_{nl}}{g_{1s}}f_{\nu'}x_{1s}- (1+f_{\nu'})x_{nl}\right]
\frac{d\Lambda_{nl}}{d\nu}
\label{eq:knu}
\end{equation}
and
\begin{equation}
f^0_\nu = \frac{\sum_{nl} (1+f_{\nu'})x_{nl}}
{\sum_{nl} [ (g_{nl}/g_{1s})f_{\nu'}x_{1s} - (1+f_{\nu'})x_{nl}]}.
\end{equation}
Here $\kappa_\nu$ can be thought of as an optical depth per unit frequency (this is more convenient in cosmology than optical depth per g$\,$cm$^{-2}$ 
as is usual in stellar structure), and $f^0_\nu$ is the photon phase space density that one would have if the level populations were maintained but the 
density were taken to $\infty$ (high optical depth limit).  Note that the ``optical depth'' used here is only the optical depth to two-photon 
transitions and does not include Rayleigh or Thomson scattering.

Our goal here is to develop an analytic approximation to Eq.~(\ref{eq:2b1}) valid in the red damping wing of Ly$\alpha$.  To do this we make three 
assumptions.  We begin by making the steady-state approximation that the time derivative on the left-hand side can be dropped.  Physically this 
corresponds to the statement that the radiation spectrum in the vicinity of Ly$\alpha$ changes slowly compared to the time a photon takes to redshift 
out of the optically thick part of the damping wing.  Secondly, we assume that the excited levels ($n\ge 2$) are in equilibrium with $2p$, i.e. 
$x_{nl}/x_{2p}$ is given by the Boltzmann factor $[(2l+1)/3]e^{-(E_{nl}-E_{2p})/T_{\rm r}}$.  This, combined with $x_{2p}\ll x_{1s}$, allows us to 
write
\begin{equation}
f^0_\nu \approx \frac{x_{2p}}{3x_{1s}} e^{h(\nu-\nu_{{\rm Ly}\alpha})/T_{\rm r}}.
\label{eq:f0n1}
\end{equation}
Our third assumption is that we may approximate $d\Lambda_{nl}/d\nu$ by the single pole approximation, $d\Lambda_{nl}/d\nu\propto(\nu-\nu_{{\rm 
Ly}\alpha}){^{-2}}$, in which the two-photon differential decay rate is proportional to the inverse-square of the detuning from the Ly$\alpha$ 
resonance.  Because of the pole in the two-photon matrix element ${\cal M}$ from the $2p$ intermediate state, this is the leading-order term in the 
power series expansion of $d\Lambda_{nl}/d\nu$ around $\nu=\nu_{{\rm Ly}\alpha}$.  The coefficient of this pole is taken from Eq.~(\ref{eq:m}): in the 
vicinity of Ly$\alpha$, we have
\begin{equation}
\frac{d\Lambda_{nl}}{d\nu} \approx \frac{27\alpha_{\rm fs}^6\nu_{2p,1s}^3\nu_{nl,2p}^3
}{a_0^4{\cal R}^6(\nu-\nu_{{\rm Ly}\alpha})^2}
\left|\langle nl||{\bf r}||2p\rangle\langle 2p||{\bf r}||1s\rangle\right|^2.
\end{equation}
It follows that
\begin{equation}
\kappa_\nu \approx
\frac{\bar W}{(\nu-\nu_{{\rm Ly}\alpha})^2}
e^{-h(\nu-\nu_{{\rm Ly}\alpha})/T_{\rm r}},
\label{eq:kn1}
\end{equation}
where
\begin{eqnarray}
\bar W &\approx& \frac{c^3n_{\rm H}x_{1s}}{8\pi H\nu_{2p,1s}^3} \sum_{nl,n\ge 3} 
\frac{2l+1}{e^{h\nu_{nl,2p}/T_{\rm r}}-1}
\nonumber \\ && \times
\frac{27\alpha_{\rm fs}^6\nu_{2p,1s}^3\nu_{nl,2p}^3
}{a_0^4{\cal R}^6}
\left|\langle nl||{\bf r}||2p\rangle\langle 2p||{\bf r}||1s\rangle\right|^2.
\end{eqnarray}
[We have neglected $x_{nl}$ in Eq.~(\ref{eq:knu}) since even with the factor of $1+f_{\nu'}$ instead of $f_{\nu'}$ it will be negligible compared to 
$x_{1s}$; this is physically equivalent to neglecting two-photon decays where the higher-energy photon is stimulated.]
This expression can be reconstructed in terms of Einstein coefficients:
\begin{equation}
\bar W\approx \frac{c^3n_{\rm H}A_{2p,1s}x_{1s}}{8\pi H\nu_{2p,1s}^3}
\sum_{nl,n\ge 3} \frac{2l+1}{e^{h\nu_{nl,2p}/T_{\rm r}}-1}
\frac{A_{nl,2p}}{4\pi^2}.
\end{equation}
The prefactor is recognizable as the Ly$\alpha$ (Gunn-Peterson) optical depth,
\begin{equation}
\tau_{{\rm Ly}\alpha} \equiv \frac{3c^3n_{\rm H}A_{2p,1s}x_{1s}}{8\pi H\nu_{2p,1s}^3},
\label{eq:gp}
\end{equation}
so we write
\begin{equation}
\bar W\approx \frac{\tau_{{\rm Ly}\alpha}}{4\pi^2}
\sum_{nl,n\ge 3}\frac{2l+1}3 \frac{A_{nl,2p}}{e^{h\nu_{nl,2p}/T_{\rm r}}-1}.
\label{eq:w}
\end{equation}

With this set of approximations, the radiative transfer equation reduces to
\begin{equation}
\frac{df_\nu}{d\nu} = \frac {\bar W}{(\nu-\nu_{{\rm Ly}\alpha})^2}
\left[ e^{h(\nu-\nu_{{\rm Ly}\alpha})/T_{\rm r}}f_\nu - \frac{x_{2p}}{3x_{1s}} \right].
\label{eq:2b1.9}
\end{equation}
The change of variables $y=h(\nu-\nu_{{\rm Ly}\alpha})/T_{\rm r}$ and $f_\nu=x_{2p}\Phi(y)/(3x_{1s})$ gives
\begin{equation}
\frac{d\Phi}{dy} = \frac{W}{y^2}(e^y\Phi-1),
\label{eq:2b2}
\end{equation}
where $W\equiv h{\bar W}/T_{\rm r}$ is a dimensionless number that quantifies the normalization of the two-photon opacity.  We can understand its 
physical significance from Eq.~(\ref{eq:kn1}).  For $W\ll 1$ the integrated two-photon optical depth at frequencies less than some critical value 
$\nu_{\rm 
c}<\nu_{{\rm Ly}\alpha}$ is
\begin{equation}
\tau_< = \int^{\nu_{\rm c}} \kappa_\nu\,d\nu
\approx \frac{T_{\rm r}}h \frac W{\nu_{{\rm Ly}\alpha}-\nu_{\rm c}},
\end{equation}
so two-photon absorption is optically thick out to a detuning $\sim WT_{\rm r}/h$ from the Ly$\alpha$ line and in the optically thick region the 
exponential factor in Eq.~(\ref{eq:2b1.9}) is a perturbation.  The physical situation realized in recombination is always $W<1$ -- we find a maximum of 
$W=0.034$ -- but the corrections due to non-negligible $W$ 
are important for the accuracy desired in upcoming CMB experiments.  

\begin{figure}
\includegraphics[angle=-90,width=3.2in]{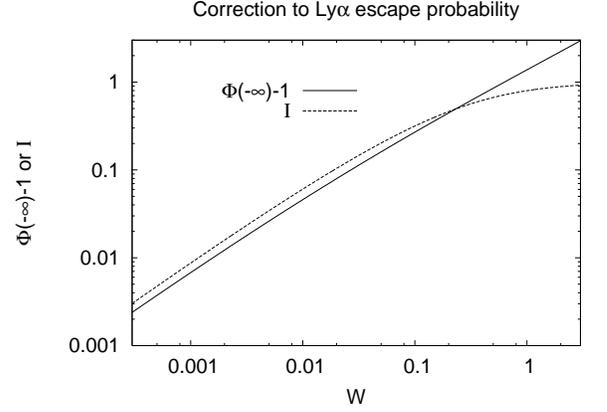}
\caption{\label{fig:phi-inf}{\em Solid line:} The correction to the Ly$\alpha$ escape probability, $\Phi(-\infty)$, as a function of $W$.
{\em Dashed line:} The integral ${\cal I}$ that describes the number of photons on the blue side 
of Ly$\alpha$ emitted in two-photon decays.
Note that as $W\rightarrow 0$ both $\Phi(-\infty)-1$ and ${\cal I}\rightarrow 0$.}
\end{figure}

Equation~(\ref{eq:2b2}) forces the boundary condition 
$\Phi(y=0)=1$, and the equation can be evolved by integrating to the left (negative $y$).  It is readily seen that as $W\rightarrow 0$ this boundary 
condition at $\Phi(y=0)$ forces $\Phi(-\infty)\rightarrow 1$.
The numerical integration of Eq.~(\ref{eq:2b2}) with an implicit ODE solver is extremely fast compared to the 
rate coefficient evaluations and linear algebra required when solving recombination, so we have simply solved Eq.~(\ref{eq:2b2}) each time a 
value of $\Phi(-\infty)$ is needed.  [Since $\Phi(-\infty)$ depends only on the single independent variable $W$ one could generate a table of values 
and use spline interpolation if the need arose for a faster evaluation.]  The numerical solution is shown in Fig.~\ref{fig:phi-inf}.

The net rate of two-photon transitions to $1s$ that involve photons in the red tail of Ly$\alpha$ can be obtained as follows.  We compare 
Eq.~(\ref{eq:xnl}), which gives the net $nl\rightarrow 1s$ rate via these transitions, to Eq.~(\ref{eq:dfnu}), which gives the rate of production of 
photons in the red tail of Ly$\alpha$.  Integrating the latter from frequency $\nu_1$ to $\nu_{{\rm Ly}\alpha}$, and making the steady state 
approximation again to remove $\dot f_\nu$, gives
\begin{equation}
0 = H\int_{\nu_1}^{\nu_{{\rm Ly}\alpha}} \nu\frac{\partial f_\nu}{\partial\nu}d\nu
+ \sum_{nl} \int_{\nu_1}^{\nu_{{\rm Ly}\alpha}} \frac{c^3n_{\rm H}}{8\pi\nu^2}\Delta_{nl}(\nu)d\nu.
\end{equation}
Now if almost all of the two-photon decays from the $n\ge 3$ levels emit one of the photons just below Ly$\alpha$, we may take $\nu\approx\nu_{{\rm 
Ly}\alpha}$ in the prefactor of the first integral and the denominator of the second integral.  Then
\begin{equation}
0 = H\nu_{{\rm Ly}\alpha}(f_{\nu_{{\rm Ly}\alpha}}-f_{\nu_1}) + \frac{c^3n_{\rm H}}{8\pi\nu_{{\rm Ly}\alpha}^2}
\sum_{nl} \int_{\nu_1}^{\nu_{{\rm Ly}\alpha}} \Delta_{nl}(\nu)d\nu.
\end{equation}
The last integral is the net rate of two-photon decays with the higher-energy photon emitted between $\nu_1$ and Ly$\alpha$.  (Here ``net'' means with 
the two-photon absorptions subtracted.)  Since $\Delta_{nl}(\nu)$ is strongly peaked near Ly$\alpha$ due to the shape of the two-photon spectrum, we 
may take $\nu_1$ to be redward of the significant two-photon emission, equate the last integral with $\dot x_{nl}|_{2\gamma,\nu<\nu_{{\rm Ly}\alpha}}$, 
and then have
\begin{equation}
0 = H\nu_{{\rm Ly}\alpha}(f_{\nu_{{\rm Ly}\alpha}}-f_{\nu_1}) + \frac{c^3n_{\rm H}}{8\pi\nu_{{\rm Ly}\alpha}^2}
\Delta \dot x_\downarrow^{\rm(B)},
\end{equation}
where $\Delta x_\downarrow^{\rm(B)}$ is the net rate of two-photon transitions to $1s$ that involve photons in the red tail of Ly$\alpha$.  Recalling 
the definition of $\Phi(y)$, that $\nu_{{\rm Ly}\alpha}$ and $\nu_1$ correspond to $y=0$ and $y=-\infty$ respectively, and that $\Phi(0)=1$, we find
\begin{equation}
\Delta\dot x_\downarrow^{\rm(B)} = \frac{8\pi H\nu_{{\rm Ly}\alpha}^3}{c^3n_{\rm H}}
\;\frac{x_{2p}}{3x_{1s}}[\Phi(-\infty)-1].
\label{eq:xdown-b1}
\end{equation}
This analytical approximation is not a surprise: it is just an extension of the Peebles \cite{1968ApJ...153....1P} argument that the net decay rate via 
emission/absorption of photons in a narrow region of the electromagnetic spectrum that is taken to be in steady state is equal to the difference in 
phase space densities on the two sides of the region, multiplied by the appropriate normalization factors.  In fact, Eq.~(\ref{eq:xdown-b1}) without 
the factor $\Phi(-\infty)-1$ (and with a correction for blackbody photons entering on the blue side) {\em is} the Peebles formula for the Ly$\alpha$ 
decay rate, so we can think of $\Phi(-\infty)$ as representing an effective 
enhancement of the Ly$\alpha$ production rate due to two-photon transitions into the red damping wing.  In our approximation this enhancement factor 
depends only on $W$.

\begin{figure}
\includegraphics[angle=-90,width=3.2in]{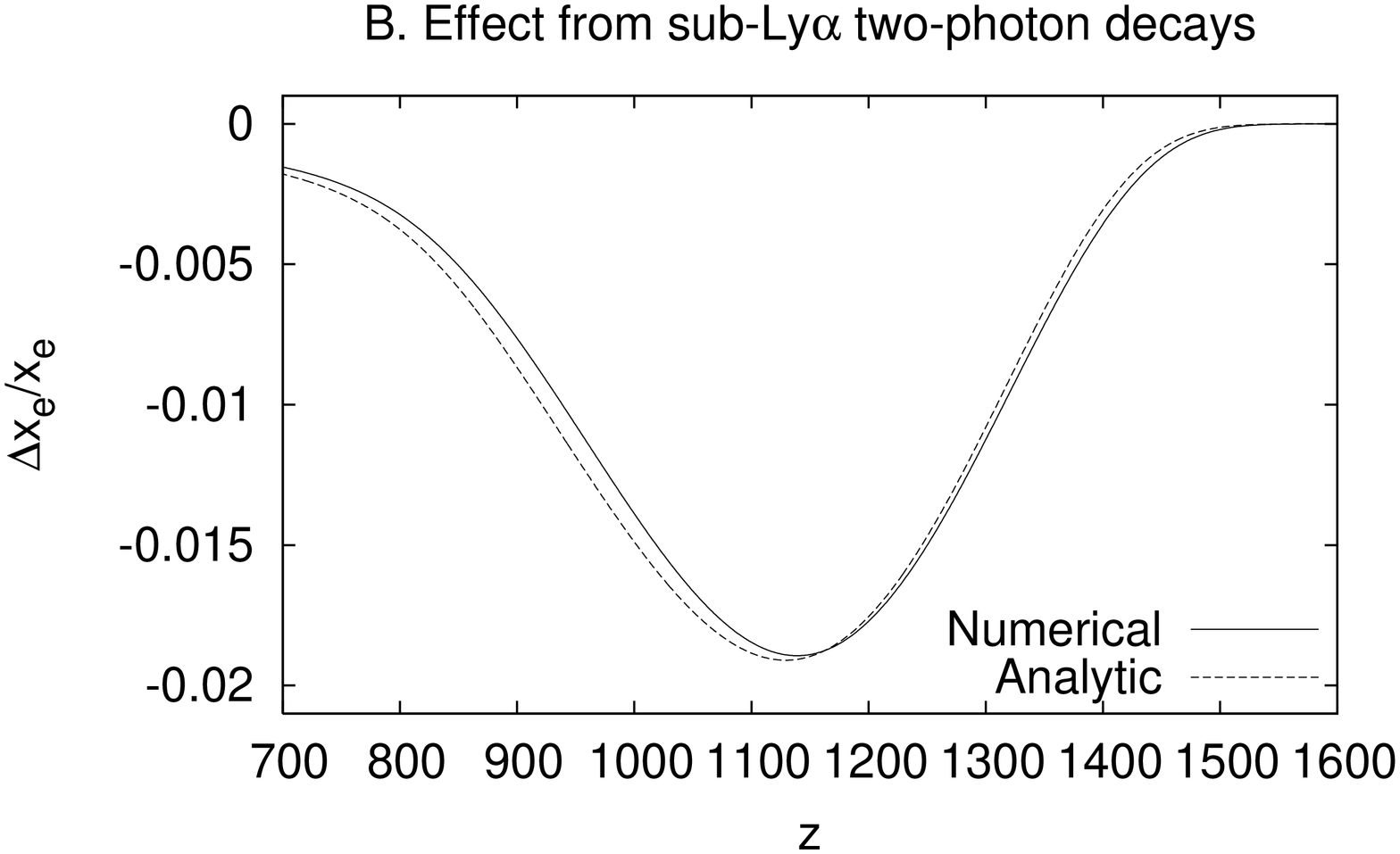}
\caption{\label{fig:diffb}A comparison of methods for handling the \subla\ $n\ge 3$ two-photon decays.
We show the numerical radiative transfer method (solid line) and the analytic approximation of Eq.~(\ref{eq:analytic-b}) (dashed line).}
\end{figure}

We now have all of the pieces required to incorporate the additional decay rate given by Eq.~(\ref{eq:xdown-b1}) into our recombination model.
Rewriting the prefactor of Eq.~(\ref{eq:xdown-b1}) in terms of $\tau_{{\rm Ly}\alpha}$, we can simplify 
Eq.~(\ref{eq:xdown-b1}) to
\begin{equation}
\Delta\dot x_\downarrow^{\rm(B)} = \frac{A_{2p,1s}}{\tau_{{\rm Ly}\alpha}} x_{2p}
[\Phi(-\infty)-1],
\label{eq:analytic-b}
\end{equation}
where $\Phi(-\infty)$ is evaluated as a function of $W$.  Note that in the $W\rightarrow 0$ limit, corresponding to negligible 
two-photon opacity, $\Phi(-\infty)\rightarrow 1$ and this correction vanishes.
We have compared Eq.~(\ref{eq:analytic-b}) to the full numerical result for the $n\ge 3$ two-photon transitions with $\nu<\nu_{{\rm Ly}\alpha}$ in 
Fig.~\ref{fig:diffb}.

\subsection{\Supla\ two-photon decays}
\label{ss:c}

We now turn to the \supla\ two-photon decays from $n\ge 3$ levels of \HI.
These decays are fundamentally different from the \subla\ decays considered in the previous section.  The reason is that 
a photon emitted at frequencies below Ly$\alpha$ has a nonzero probability of escaping the Ly$\alpha$ line, after which it is extremely unlikely to be 
re-absorbed; it may be Thomson or Rayleigh scattered, but it is unlikely to re-excite an \HI\ atom and will consequently have little further impact on 
recombination.  In contrast, all photons emitted above the Ly$\alpha$ frequency will eventually redshift down to Ly$\alpha$ and be absorbed.  That is, 
for every \supla\ two-photon decay there is an additional Ly$\alpha$ absorption, so the implied net number of decays to the 
\HI\ $1s$ state is zero.  For this reason, CS08 did not include these decays in their analysis.  However, we can see from Fig.~\ref{fig:mods} that 
these decays do indeed result in a change in the recombination history.  This is because of the {\em delay} between the two-photon decay that produces 
the photon at $\nu>\nu_{{\rm Ly}\alpha}$ and the Ly$\alpha$ absorption.  This section constructs an analytic approximation that accounts for the effect 
of this delay on recombination.

The key to our analytic approximation is to consider $x_+^{2\gamma}$, the number of non-thermal photons above the Ly$\alpha$ frequency per H atom 
produced by the two-photon decays.  Initially the photon spectrum is thermal, so $x_+^{2\gamma}=0$.  Long after recombination is over and all of the 
photons have redshifted below Ly$\alpha$, we also have $x_+^{2\gamma}=0$.  The net rate of decays to the ground state due to two-photon decays with 
$\nu>\nu_{{\rm Ly}\alpha}$ and re-absorption of the emitted photons is
\begin{equation}
\dot x_\downarrow^{\rm(C)} = -\dot x_+^{2\gamma}.
\label{eq:analytic-c}
\end{equation}
The total number of such decays per H atom is
\begin{equation}
\int \dot x_\downarrow^{\rm(C)}\,dt = -\int\dot x_+^{2\gamma}\,dt = x_+^{2\gamma}(t_{\rm final}) - x_+^{2\gamma}(t_{\rm init}) = 0,
\end{equation}
as expected.

It only remains to find $x_+^{2\gamma}(t)$.  By mode counting, we find
\begin{equation}
x_+^{2\gamma} = \frac{8\pi}{c^3n_{\rm H}}\int_{\nu_{{\rm Ly}\alpha}}^\infty \nu^2 \Delta f_\nu\,d\nu,
\label{eq:xp2}
\end{equation}
where $\Delta f_\nu$ is the non-thermal phase space density of photons.  Switching to the variable $y=h(\nu-\nu_{{\rm Ly}\alpha})/T_{\rm r}$, and 
taking the approximation that the distortion photons above Ly$\alpha$ are mostly in the regime $\nu/\nu_{{\rm Ly}\alpha}-1\ll 1$,
\begin{equation}
x_+^{2\gamma} \approx \frac{8\pi\nu_{{\rm Ly}\alpha}^2T_r}{c^3n_{\rm H}h}\int_0^\infty \Delta f_\nu\,dy.
\label{eq:xplus}
\end{equation}
We may estimate the spectral distortion using Eq.~(\ref{eq:2b1}).  Rewriting it in terms of the distortion $\Delta f_\nu$, we find
\begin{equation}
\frac{\Delta\dot f_\nu}{H\nu} = \frac{\partial \Delta f_\nu}{\partial\nu} - \kappa_\nu (\Delta f_\nu - \Delta f^0_\nu),
\label{eq:2c1}
\end{equation}
where
\begin{equation}
\Delta f^0_\nu=f_0^\nu - \frac 1{e^{h\nu/T_{\rm r}}-1}
\end{equation}
and the optical depth per unit frequency $\kappa_\nu$ is the same as that from Sec.~\ref{ss:b}.
Under the same assumptions from Sec.~\ref{ss:b} that led to Eq.~(\ref{eq:f0n1}), we find
\begin{equation}
\Delta f^0_\nu \approx \left(\frac{x_{2p}}{3x_{1s}} - e^{-h\nu_{{\rm Ly}\alpha}/T_{\rm r}}\right)
 e^{-h(\nu-\nu_{{\rm Ly}\alpha})/T_{\rm r}}.
\end{equation}
We now assume the distortion is in steady state, i.e. that the left-hand side of Eq.~(\ref{eq:2c1}) is zero.
Defining
\begin{equation}
\Psi(\nu) = \left(\frac{x_{2p}}{3x_{1s}} - e^{-h\nu_{{\rm Ly}\alpha}/T_{\rm r}}\right)^{-1}\Delta f_\nu,
\label{eq:psidef}
\end{equation}
we convert Eq.~(\ref{eq:2c1}) into
\begin{equation}
0 = \frac{d\Psi}{d\nu} - \kappa_\nu(\Psi-e^{-y}).
\label{eq:2c1.5}
\end{equation}
Using Eq.~(\ref{eq:kn1}) for $\kappa_\nu$ we convert this into
\begin{equation}
\frac{d\Psi}{dy} = W(e^y\Psi-1).
\label{eq:2c2}
\end{equation}
This is the same equation as Eq.~(\ref{eq:2b2}), except that now the interesting region is $0<y<\infty$, and the ``initial condition'' is that 
$\Psi(\infty)=0$.  We define the integral
\begin{equation}
{\cal I} = \int_0^\infty \Psi\,dy,
\end{equation}
which is a function of $W$.  Then Eq.~(\ref{eq:xplus}) can be re-written as
\begin{equation}
x_+^{2\gamma} \approx \frac{8\pi\nu_{{\rm Ly}\alpha}^2T_r}{c^3n_{\rm H}h}
\left(\frac{x_{2p}}{3x_{1s}} - e^{-h\nu_{{\rm Ly}\alpha}/T_{\rm r}}\right){\cal I}.
\label{eq:xplus2}
\end{equation}

The integral ${\cal I}$ as a function of $W$ is plotted in Fig.~\ref{fig:phi-inf}.  We evaluate it with an implicit ODE each time it is needed.  From 
Eq.~(\ref{eq:2c2}) we see that one should have ${\cal I}\rightarrow 0$ as $W\rightarrow 0$, and $\Psi\rightarrow e^{-y}$, ${\cal I}\rightarrow 1$ as 
$W\rightarrow \infty$.  Figure~\ref{fig:phi-inf} shows that this is precisely what happens.

Equation~(\ref{eq:analytic-c}) presents an implementation difficulty: it depends on the derivative of $x_+^{2\gamma}$, which itself depends on $\dot 
x_e$ because of the implicit dependence of $x_{1s}$ and $x_{2p}$ in Eq.~(\ref{eq:xplus2}) on $x_e$.  In principle one could solve this by treating 
Eq.~(\ref{eq:analytic-c}) as an implicit equation for $\dot x_\downarrow^{\rm(C)}$.  In practice, however, the effect of the two-photon decays from 
highly excited states is only a small perturbation on the recombination problem, so we have calculated Eq.~(\ref{eq:xplus2}) using the ``standard'' 
recombination history described in Sec.~\ref{ss:mla}.  We then calculate $\dot x_+^{2\gamma}$ for use in Eq.~(\ref{eq:analytic-c}) by numerically 
differentiating this precomputed table of $\dot x_+^{2\gamma}(t)$.

\begin{figure}
\includegraphics[angle=-90,width=3.2in]{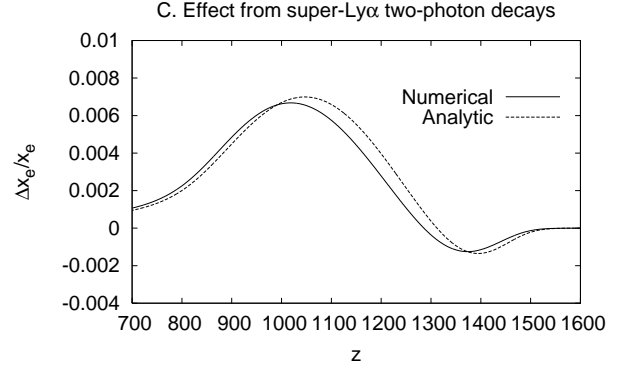}
\caption{\label{fig:diffc}A comparison of methods for handling the \supla\ $n\ge 3$ two-photon decays.
We show the numerical radiative transfer method (solid line) and the analytic approximation of Eqs.~(\ref{eq:analytic-c}) and (\ref{eq:xplus2})
(dashed line).}
\end{figure}

The result is shown in Fig.~\ref{fig:diffc}.  One can see that the analytic approximation described here not only captures the qualitative fact that 
recombination is first accelerated and then delayed, because $x_+^{2\gamma}$ first increases and then decreases, but also captures the quantitative 
magnitude of the effect.  The maximum of $x_+^{2\gamma}$ is 0.0061 at $z=1360$, and one can see from Fig.~\ref{fig:diffc} that this is indeed when the 
two-photon decays with $\nu>\nu_{{\rm Ly}\alpha}$ switch from speeding up recombination to slowing it down.

\begin{figure}
\includegraphics[angle=-90,width=3.2in]{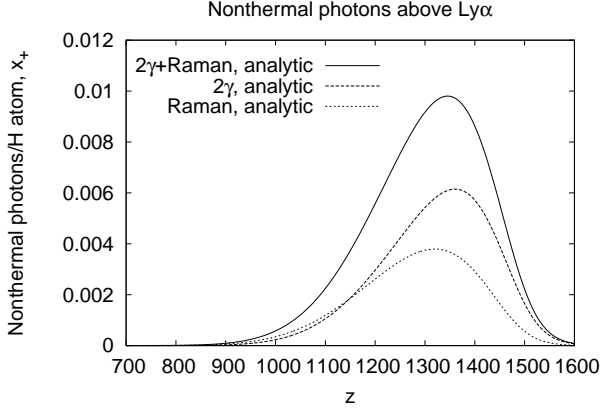}
\caption{\label{fig:xplus}The number of nonthermal photons above Ly$\alpha$ per H atom.  The long-dashed curve shows the analytic 
estimate for $x_+^{2\gamma}$ (Sec.~\ref{ss:c}); the short-dashed curve shows the analytic estimate for $x_+^{\rm R}$ (Sec.~\ref{ss:d}); and the solid 
curve shows their sum $x_+^{2\gamma}+x_+^{\rm R}$.}
\end{figure}

The abundance $x_+^{2\gamma}$ of nonthermal photons above Ly$\alpha$ from two-photon decay using the analytic estimate is shown in 
Fig.~\ref{fig:xplus}.

\subsection{Raman scattering}
\label{ss:d}

We now consider Raman scattering.  Selection rules imply that Raman scattering to the $1s$ level is only possible from $s$ and $d$ levels.  The most
populated such level is $2s$, so we expect that Raman scattering is dominated by the $2s\leftrightarrow 1s$ process.  This process contains
contributions from two types of photons: the photons near the peak of the blackbody spectrum, and the photons near H$\alpha$ which contribute resonant
Raman scattering,
\begin{equation}
{\rm H}(2s) + h\nu' \rightarrow {\rm H}(3p) \rightarrow {\rm H}(1s) + h\nu,
\label{eq:3p}
\end{equation}
which is resonant when $\nu'\approx \nu_{{\rm H}\alpha}$ and $\nu\approx \nu_{{\rm Ly}\beta}$.
These pieces are (mostly) distinct: the radiation temperature is typically $0.02h{\cal R}$ ($z=1200$) whereas the H$\alpha$ energy 
is $0.14h{\cal R}$.  Thus the rate of Raman scattering as a function of frequency $\nu'$ of the incoming photon has two peaks -- a nonresonant peak due 
to the large abundance of photons with energies of a few times $T_{\rm r}$, and a resonant peak at H$\alpha$.  There are
additional peaks due to the other Balmer resonances H$\beta$, H$\gamma$, etc., but the abundance of these photons is at least a factor of a few less 
than H$\alpha$ (due to the Boltzmann factor) and the dipole matrix elements are smaller.  This behavior can be seen in Fig.~\ref{fig:rates}.

The two sub-classes of Raman scattering transitions -- nonresonant and resonant (H$\alpha$) -- are considered separately here.

\subsubsection{Nonresonant scattering}

In terms of effects on
recombination, the nonresonant Raman scattering is similar to the \supla\ two-photon decays in that a Raman scattering that 
produces an \HI\
atom in the $1s$ state also produces a photon above the Ly$\alpha$ frequency.  This photon will eventually redshift into the Ly$\alpha$ line and excite
a hydrogen atom, so again the net number of decays to the \HI\ $1s$ state is zero.  Again the importance for recombination derives from the delay
between emission of the photon and its re-absorption.  Thus Eq.~(\ref{eq:analytic-c}) applies to Raman scattering as well, except that in the case of
Raman scattering we will need a formula for $x_+^{\rm R}$, the number of non-thermal photons above Ly$\alpha$ contributed by nonresonant Raman 
scattering:
\begin{equation}
\dot x_\downarrow^{\rm(D1)} = -\dot x_+^{\rm R}.
\label{eq:analytic-d1}
\end{equation}

Our challenge now is to find an equation for $x_+^{\rm R}$.  Reviewing the steps that lead to Eq.~(\ref{eq:2c1.5}) shows that these are valid in the 
case of Raman scattering, except that we must replace the two-photon opacity with the Raman opacity,
\begin{equation}
\kappa_\nu = \frac{c^3n_{\rm H}}{8\pi H\nu^3}\sum_{nl}\left[ \frac{g_{nl}}{g_{1s}}(1+f_{\nu'})x_{1s} - f_{\nu'}x_{2s} \right] \frac{dK_{nl}}{d\nu},
\label{eq:knu1}
\end{equation}
in place of Eq.~(\ref{eq:knu}).  Taking the $2s$ term to be dominant, recalling that $x_{2s}\ll x_{1s}$, and using the blackbody formula for 
$f_{\nu'}$, we can re-write this as
\begin{equation}
\kappa_\nu \approx \frac{c^3n_{\rm H}}{8\pi H\nu^3} \frac{x_{1s}}{1-e^{-y}} \frac{dK_{2s}}{d\nu},
\label{eq:knu-r}
\end{equation}
where $y=h\nu'/T_{\rm r}=h(\nu-\nu_{{\rm Ly}\alpha})/T_{\rm r}$.

The Raman scattering rate coefficient $dK_{2s}/d\nu$ can be re-written using 
Eq.~(\ref{eq:k}) by noting that so long as $\nu'\ll \nu_{{\rm H}\alpha}$, we may take $2p$ to be the dominant intermediate 
state.  Using the matrix element $\langle 2s||{\bf r}||2p\rangle^2=27a_0^2$:
\begin{equation}
\frac{dK_{2s}}{d\nu}\approx \frac{\alpha_{\rm fs}^6\nu^3\nu'}{{\cal R}^4a_0^2}\left|\langle 1s||{\bf r}||2p\rangle\right|^2.
\end{equation}
The squared matrix element is related to the Ly$\alpha$ decay rate via the dipole emission formula
\begin{equation}
A_{{\rm Ly}\alpha} = \frac29\pi\alpha_{\rm fs}^3\frac{\nu_{{\rm Ly}\alpha}^3}{{\cal R}^2a_0^2}\left|\langle 1s||{\bf r}||2p\rangle\right|^2,
\end{equation}
from which we may derive
\begin{equation}
\frac{dK_{2s}}{d\nu}\approx \frac 9{2\pi}A_{{\rm Ly}\alpha}\left(\frac{\nu}{\nu_{{\rm Ly}\alpha}}\right)^3
\alpha_{\rm fs}^3 \frac{\nu'}{{\cal R}^2}.
\end{equation}
This is just the lowest-order term in the power series expansion of $dK_{2s}/d\nu$; there are corrections of order $\nu'{^2}$, $\nu'{^3}$, etc.  Due to 
the 
large prefactor of the next-lowest term we include it:
\begin{equation}
\frac{dK_{2s}}{d\nu}\approx \frac 9{2\pi}A_{{\rm Ly}\alpha}\left(\frac{\nu}{\nu_{{\rm Ly}\alpha}}\right)^3
\alpha_{\rm fs}^3 \frac{\nu'}{{\cal R}^2}\left( 1 + 8.15\frac{\nu'}{\cal R} \right).
\end{equation}
The coefficient 8.15 was determined by numerical differentiation of $dK_{2s}/d\nu$.  It is positive because of constructive interference between the 
dominant $2s\rightarrow 2p\rightarrow 1s$ Raman scattering pathway and the neighboring $2s\rightarrow 3p\rightarrow 1s$ pathway.

Making this approximation and plugging into Eq.~(\ref{eq:knu-r}), we find
\begin{equation}
\kappa_\nu \approx \frac{9c^3n_{\rm H}A_{{\rm Ly}\alpha}}{16\pi^2 H\nu_{{\rm Ly}\alpha}^3} \frac{x_{1s}}{1-e^{-y}} 
\alpha_{\rm fs}^3 \left( 1 + 8.15\frac{\nu'}{\cal R} \right) \frac{\nu'}{{\cal R}^2}.
\end{equation}
We may re-write this using the Ly$\alpha$ optical depth, Eq.~(\ref{eq:gp}), as
\begin{equation}
\kappa_\nu \approx \frac 3{2\pi}\alpha_{\rm fs}^3 \tau_{{\rm Ly}\alpha}
\left(\frac{T_{\rm r}}{h\cal R}\right)^2 \left( 1 + 8.15\frac{T_{\rm r}}{h{\cal R}}y \right)
\frac{\nu'}{1-e^{-y}}.
\end{equation}
The Raman scattering version of Eq.~(\ref{eq:2c1.5}) is then
\begin{equation}
\frac{d\Psi}{dy} = V\left( 1 + 8.15\frac{T_{\rm r}}{h{\cal R}}y
\right)\frac y{1-e^{-y}}(\Psi-e^{-y}),
\label{eq:psi-y}
\end{equation}
where
\begin{equation}
V = \frac 3{2\pi}\alpha_{\rm fs}^3 \tau_{{\rm Ly}\alpha}
\left(\frac{T_{\rm r}}{h\cal R}\right)^2.
\label{eq:v-def}
\end{equation}

Equation~(\ref{eq:psi-y}) has the integral solution,
\begin{equation}
\Psi(y) = V\int_y^\infty \frac{y'e^{-y'}}{1-e^{-y'}} \left(1 + 8.15\frac{T_{\rm r}}{h{\cal R}}y'\right)
e^{-\tau(y,y')} dy',
\label{eq:integralsolution}
\end{equation}
where
\begin{equation}
\tau(y,y') = V\int_y^{y'} \frac {y''}{1-e^{-y''}} \left(1 + 8.15\frac{T_{\rm r}}{h{\cal R}}y''\right)\,dy''
\end{equation}
is the optical depth between $y$ and $y'$.
We may then write the dimensionless integral of the spectral distortion:
\begin{eqnarray}
{\cal J} &\equiv& \int_0^\infty \Psi(y)\,dy
\nonumber\\
&=& V\int_0^\infty dy' \, \frac{y'e^{-y'}}{1-e^{-y'}}
\nonumber\\
&& \times \left(1 + 8.15\frac{T_{\rm r}}{h{\cal R}}y'\right)
\int_0^{y'} dy\, e^{-\tau(y,y')}.
\end{eqnarray}
If $\tau$ were negligible then the inner integral would collapse to give $y'$ and the outer integral would be dominated by the region $y'\sim 2$.  In 
practice for $V\ll 1$ (the maximum is 0.046 at $z=1184$) $\tau$ will be $\ll 1$ in this region; for much larger $y'$ it may be significant but all it 
does is suppress a portion of the integral that is already negligible.  Thus we may neglect $\tau$ and replace the integral $\int_0^{y'} dy\, 
e^{-\tau(y,y')}$ with $y'$.  This gives
\begin{eqnarray}
{\cal J} &\approx& V\int_0^\infty dy' \, \frac{y'{^2}e^{-y'}}{1-e^{-y'}} \left(1 + 8.15\frac{T_{\rm r}}{h{\cal R}}y'\right)
\nonumber\\
&=& \left[2\zeta(3) + 8.15\frac{T_{\rm r}}{h{\cal R}}\frac{\pi^4}{15} \right]V
\nonumber\\
&=& 2\zeta(3) V \left( 1 + 22.0\frac{T_{\rm r}}{h{\cal R}}\right).
\label{eq:cal-j}
\end{eqnarray}
We may now assemble all of the pieces and plug them into Eq.~(\ref{eq:xp2}): the spectral distortion is given in terms of $\Psi$ by 
Eq.~(\ref{eq:psidef}), and the integral of $\Psi$ is Eq.~(\ref{eq:cal-j}).  The normalization $V$ is given by Eq.~(\ref{eq:v-def}).  Combining 
everything gives
\begin{eqnarray}
x_+^{\rm R} &=& \frac{27}2\zeta(3)\left(\frac{T_{\rm r}}{hc}\right)^3 \frac{\tau_{{\rm Ly}\alpha}}{n_{\rm H}}
\left( 1 + 22.0\frac{T_{\rm r}}{h{\cal R}}\right)
\nonumber\\ && \times
\left( \frac{x_{2p}}{3x_{1s}} - e^{-h\nu_{{\rm Ly}\alpha}/T_{\rm r}} \right).
\label{eq:xpr}
\end{eqnarray}

The same implementation difficulty arises with Eq.~(\ref{eq:analytic-d1}) as arose with Eq.~(\ref{eq:analytic-c}), namely that in order to compute 
$\Delta \dot x_\downarrow^{\rm(D)}$ one needs to know $\dot x_+^{\rm R}$ which in turn requires one to already know the recombination history.  We 
solve this problem by taking $x_+^{\rm R}$ from the standard recombination history (Sec.~\ref{ss:mla}), just as we did in Sec.~\ref{ss:c}.

\subsubsection{${\rm H}\alpha$ resonance}

Resonant Raman scattering via the $3p$ intermediate state [Eq.~(\ref{eq:3p})] will induce a red damping wing correction to the Ly$\beta$ escape 
probability similar to the corrections discussed in Sec.~\ref{ss:b} for the two-photon corrections to Ly$\alpha$.
Precisely the same calculation 
applies here except that the line width, Eq.~(\ref{eq:w}) gets replaced by
\begin{equation}
\bar W_\beta\approx \frac{\tau_{{\rm Ly}\beta}}{4\pi^2}\;
\frac{A_{3p,2s}}{1-e^{-h\nu_{3p,2s}/T_{\rm r}}}.
\label{eq:w-beta}
\end{equation}
[The only differences in the calculation are the replacement of the initial state, $2s$ instead of $nl$; the intermediate state, $3p$ instead of $2p$; 
the use of the Raman scattering rate coefficient $dK/d\nu$ in place of $d\Lambda/d\nu$; since the low-frequency photon is in the initial state the 
use of $1+f_{\nu'}$ in Eq.~(\ref{eq:knu1}) instead of $f_{\nu'}$ which appears in Eq.~(\ref{eq:knu}); and no factor of $(2l+1)/3$ because the 
Einstein coefficient $A_{3p,2s}$ in Eq.~(\ref{eq:w-beta}) has the states reversed from $A_{nl,2p}$ in Eq.~(\ref{eq:w}).]  Once again we can construct a 
dimensionless variable $W_\beta=h\bar W_\beta/T_{\rm r}$ and introduce a correction
\begin{equation}
\Delta\dot x_\downarrow^{\rm(D2)} = \frac{A_{3p,1s}}{\tau_{{\rm Ly}\beta}} x_{3p}
[\Phi(-\infty)-1],
\label{eq:analytic-d2}
\end{equation}
where $\Phi(-\infty)$ is evaluated from Eq.~(\ref{eq:2b2}) using $W_\beta$ instead of $W$.

The parameter $W_\beta$ measures the normalization of the Raman scattering opacity [to the Ly$\beta$ photon, not the H$\alpha$ photon; i.e. for the 
reverse reaction of Eq.~(\ref{eq:3p})], just as $W$ measured the normalization of the two-photon opacity to photons near Ly$\alpha$.
We find that $W_\beta$ reaches a maximum of 0.94 at 
$z=981$.  It is much larger than the corresponding $W$ for the Ly$\alpha$ line because two-photon absorption of a photon near Ly$\alpha$ requires a 
three-particle interaction between the Ly$\alpha$ photon, an atom, and a rare Balmer photon in the Wien tail of the CMB; whereas 
a photon near Ly$\beta$ can be Raman-scattered in a two-body interaction without the help of any initial-state Balmer photons.  The larger $W_\beta$ 
means that the corrections to Ly$\beta$ are of similar overall importance to the corrections to Ly$\alpha$, even though in an absolute sense Ly$\beta$ 
is less important.

The overall change in the downward decay rate from Raman scattering is the sum of the nonresonant and resonant contributions:
\begin{equation}
\Delta\dot x_\downarrow^{\rm(D)} =
\Delta\dot x_\downarrow^{\rm(D1)} +
\Delta\dot x_\downarrow^{\rm(D2)}.
\label{eq:analytic-d}
\end{equation}
However it should be noted that the downward rate D2 (Eq.~\ref{eq:analytic-d2}) is emitted in the red damping wing of Ly$\beta$ and is added to the 
Ly$\beta$ spectral distortion $f_{{\rm Ly}\beta-} \equiv f_{3p,1s-}$ appearing on the left hand side of Eq.~(\ref{eq:f-}).  It is also included as an 
increase in the $3p\rightarrow 1s$ transition rate $\dot x_{3p\rightarrow 1s}$.  It therefore feeds back at 
later times on Ly$\alpha$.  In contrast the rate D1 has all radiative transfer effects already included, and it is included along with 
Eq.~(\ref{eq:analytic-c}) in the $2p\rightarrow 1s$ rate.  (It is not obvious whether it is better to analytically approximate D1 as being a net rate 
from the $2s$ or $2p$ state, since the Raman scattering occurs from $2s$ but re-excitations are to $2p$.  In practice $2s$ and $2p$ are in equilibrium 
in the redshift range considered so this choice does not matter.)

\begin{figure}
\includegraphics[angle=-90,width=3.2in]{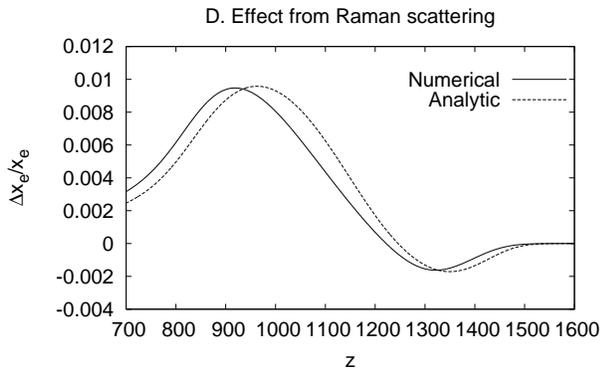}
\caption{\label{fig:diffd}A comparison of methods for handling Raman scattering.
We show the numerical radiative transfer method (solid line) and the analytic approximation of Eq.~(\ref{eq:analytic-d}) and (\ref{eq:xpr}) (dashed 
line).}
\end{figure}

The effect of this correction on recombination is shown in Fig.~\ref{fig:diffd}.  Our analytic approximation reproduces most of the features of the 
full numerical result.  The main deficiency is that the effects on $\Delta x_e/x_e$ occur earlier in the analytic calculation than in the numerical 
result by about $\Delta z\sim 50$.  This error might be due to the approximation that $x_+^{\rm R}$ instantaneously reaches its steady-state value, 
but we defer investigation of this possibility to future work.

\begin{figure}
\includegraphics[angle=-90,width=3.2in]{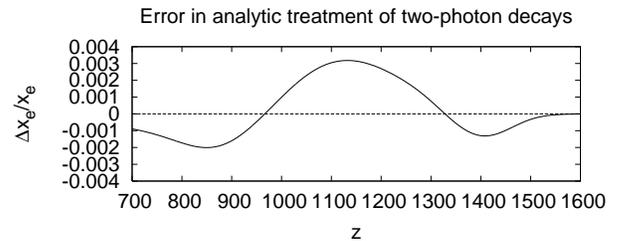}
\caption{\label{fig:analytic}The error in using analytic approximations for the two-photon effects.  The vertical axis plots $\Delta x_e/x_e$, with the 
sign convention positive where the analytic approximation overestimates $x_e$ and negative where it underestimates $x_e$.}
\end{figure}

As a summary, we show in Fig.~\ref{fig:analytic} the total error in using the analytical approximations presented here in place of more accurate 
numerical two-photon radiative transfer.  The analytic approximations have reduced the maximal fractional error in $x_e$ from 1.3\% with no treatment 
of the two-photon effects to 0.3\%, the maximal difference between the analytic and numerical treatments.

\section{Effect on the CMB}
\label{sec:cmb}

\begin{figure}
\includegraphics[angle=-90,width=3.2in]{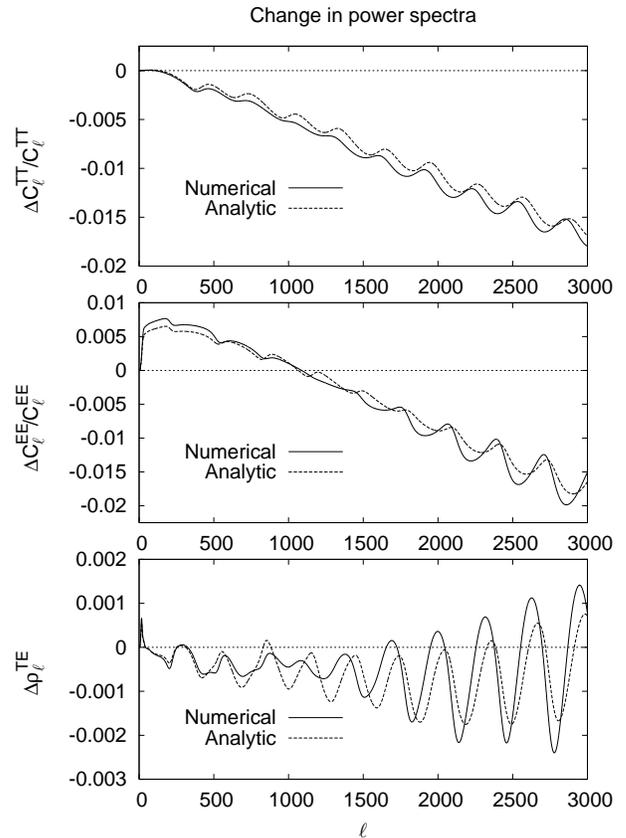}
\caption{\label{fig:change}The change in CMB temperature power spectrum, $E$-mode polarization, and cross correlation due to the two-photon processes
described in this paper.  The full calculation is shown with the solid lines, and the prediction from our simple analytic model is shown with the 
dashed lines.}
\end{figure}

We have calculated the effect of our corrected recombination history on the CMB anisotropies using the {\sc CMBFast} \cite{1996ApJ...469..437S} 
Boltzmann code.  The results are shown in Fig.~\ref{fig:change}.  For the temperature and $E$-mode polarization, we show the fractional change in the 
power spectrum $C_\ell^{TT}$ or $C_\ell^{EE}$.  For the cross-spectrum we show the change in correlation coefficient, defined by
$\rho^{TE}_\ell = C_\ell^{TE}/(C_\ell^{TT}C_\ell^{EE})^{1/2}$.
This is more useful than $\Delta 
C^{TE}_\ell/C^{TE}_\ell$, which is ill-behaved when $C^{TE}_\ell$ passes through zero.  For comparison we show the analytic 
approximations of Sec.~\ref{sec:semi}.  (Note that we denote the CMB multipoles by $\ell$ to avoid confusion with atomic orbital angular momentum $l$.)

While the mapping from the recombination history to the CMB power spectrum is in general complicated \cite{2006MNRAS.373..561L, 2008MNRAS.386.1023W}, 
the qualitative features of Fig.~\ref{fig:change} 
are easily understood.  The major effect on the temperature is that the decrease in free electron abundance in the $1100<z<1500$ range increases the 
mean free path of the photons and hence the Silk damping length \cite{1968ApJ...151..459S, 1983MNRAS.202.1169K, 1995ApJ...444..489H, 
1995PhRvD..52.3276Z}.  This means that the oscillations of the baryon-photon fluid are damped more effectively when two-photon transitions are included 
than in the standard picture, and hence the power spectrum is decreased.  This effect is most significant at the higher multipoles where damping is 
most effective.  In principle, changes in the recombination history could also have moved the surface of last scatter, thereby shifting the acoustic 
peak positions and leading to oscillations in $\Delta C_\ell^{TT}/C_\ell^{TT}$; however as one can see from the figure, this is a small effect.  This 
is 
because the accelerating effect on recombination of \subla\ two-photon decays and the decelerating effect of \supla\ decays and Raman scattering cancel 
at $z\approx 1100$.  Thus, purely by accident, the surface of last scatter is not significantly shifted.

Silk damping also suppresses the $E$-mode polarization, and again this suppression is evident in the figure at high multipoles.  However this is not 
the only effect of the new recombination history.  In our calculation, recombination is proceeding more slowly at the surface of last scatter, i.e. 
$x_e$ is decreasing more slowly than in the standard picture, so the surface of last scatter is broader.  Since generation of polarization requires a 
sufficiently low Thomson opacity to generate a local photon quadrupole by free streaming, and sufficiently high Thomson opacity to convert this 
quadrupole into polarization, the broader surface of last scatter enhances the $E$-mode polarization \cite{1995PhRvD..52.3276Z}.  Since Silk damping 
increases rapidly as one goes to small scales, the broadening of the surface of last scatter wins at low $\ell$ ($\ell\lsim 1000$) but Silk damping 
wins at high $\ell$, leading to the characteristic ``positive, then negative'' shape seen in the middle panel of Fig.~\ref{fig:change}.  At very low 
multipoles $\ell\lsim 50$ the old and new calculations agree despite the broader surface of last scatter; this is because the large-scale polarization 
is dominated by scattering at reionization, which we have left unchanged.  However, the small changes described here are unobservable at such low 
$\ell$ 
because of cosmic variance.

The temperature-polarization cross-correlation coefficient is only slightly affected ($\lsim 0.002$) because to a first approximation the Silk damping 
and the broadening of the last scattering surface only give a wavenumber-dependent re-scaling of the $T$ and $E$.  The only recombination process that 
would substantially change $\rho_\ell^{TE}$ would be to move the surface of last scatter, and as noted above our corrections to recombination do not 
significantly move the surface of last scatter.

One can estimate the importance of the changes in the recombination history for various CMB experiments by computing
\begin{equation}
Z^2 = \sum_{\ell\ell'} F_{\ell\ell'} \Delta C_\ell^{TT} \Delta C_{\ell'}^{TT},
\label{eq:z2}
\end{equation}
where $F_{\ell\ell'}$ is the Fisher matrix for the temperature power spectra.  If all cosmological parameters were known perfectly and there were no 
systematic errors then $Z$ represents the number of sigma at which the changes in the recombination history could be detected.  In the realistic case 
in which one is trying to measure the cosmological parameters, then the maximum possible bias from using the incorrect recombination history is $Z$ 
sigma; the actual bias will depend on whether the parameter in question produces a similar or orthogonal change in the $C_\ell$'s to the change from 
the 
recombination history.
For the {\slshape WMAP} 5-year Fisher matrix $F_{\ell\ell'}$ \cite{2008arXiv0803.0586D, 2008arXiv0803.0547K} we find $Z=0.40$, i.e. a 
$0.40\sigma$ effect for 
statistical errors only.  In practice the calibration, beam, and point source uncertainties are significant in the range $200\lsim \ell\lsim 
800$ relevant for 
{\slshape WMAP} \cite{2008arXiv0803.0732H, 2008arXiv0803.0593N, 2008arXiv0803.0570H, 2008arXiv0803.0577W}; if one includes them by adding these modes 
to the inverse Fisher matrix, this is reduced to 
$0.27\sigma$.  A similar calculation is possible for the ACBAR final data release \cite{2008arXiv0801.1491R}; 
here one finds a $0.48\sigma$ effect for the Fisher matrix only, which is reduced to $0.19\sigma$ after marginalizing over the beam and calibration 
uncertainty.  Thus the corrections described here have negligible impact on the interpretation of the currently published {\slshape WMAP} and ACBAR 
data \cite{2008arXiv0801.1491R, 2008arXiv0803.0547K}. However for the upcoming {\slshape Planck} mission \cite{2006astro.ph..4069T}, if we use the data 
model 
from the Dark Energy Task Force \cite{2006astro.ph..9591A} including temperature anisotropies in the 100, 143, and 217 GHz bands, and $E$-mode 
polarization in the 143 and 217 GHz bands [including $C_\ell^{EE}$ in the sum and Fisher matrix in Eq.~(\ref{eq:z2})] we find $Z=7$, i.e. a $7\sigma$ 
effect.

We have also calculated the importance of the error using the analytic approximation, i.e. we obtain $Z^2$ from Eq.~(\ref{eq:z2}) but using the 
difference between numerical and analytic results for $\Delta C_\ell^{TT}$.  This gives $Z=0.8$ for {\slshape Planck}, implying that the analytic 
treatment of two-photon decays in this paper is sufficient to reduce the errors from $7\sigma$ to $0.8\sigma$ for {\slshape Planck}.  While this is an 
enormous improvement, one would like to do better in the future, perhaps by computing the finite lag time required for $x_+^{2\gamma}$ to reach its 
steady-state value, or (less attractively) with a fudge factor fit to the full numerical result.

In addition to the CMB anisotropies, one could also consider the global spectral distortion from recombination.
While observation of this signal will be challenging due to foregrounds \cite{2006MNRAS.367.1666W} and the sensitivity 
requirements \cite{2002ApJ...581..817F}, it has prompted a large amount of work over the years because it is a direct probe of recombination physics 
\cite{1975SvAL....1..196D, 1993ASPC...51..548R, 1998A&A...336....1B, 2006MNRAS.367.1666W, 2006MNRAS.371.1939R, 2006A&A...458L..29C, 
2007arXiv0711.0594R}.  Our analysis has shown a $\sim 1$ order of magnitude increase in the spectral distortion blueward of Ly$\alpha$; however all of 
these photons are absorbed and re-processed when they redshift into the Ly$\alpha$ line and in our calculation the effect on the observable spectral 
distortion (red side of Ly$\alpha$) is at the level of a few percent.  We have not followed the lower-frequency distortion due to Balmer, Paschen, etc. 
lines in this paper, but the excited level occupations typically change by a few percent or less and we expect the final spectral distortion due to 
these lines to change by a similar amount.  Given that the spectral distortion has not yet been detected, it seems unlikely that the corrections 
considered in this paper will be important for spectral distortion studies in the forseeable future.

\section{Conclusion}
\label{sec:conc}

In this paper we have reconsidered the physics of two-photon decays in hydrogen recombination, with particular emphasis on decays from highly excited 
levels $n\ge 3$ of \HI.  We reviewed the conceptual difficulties that have plagued earlier attempts to incorporate these transitions, particularly 
those that arise from sequential ``1+1'' decays such as $3d\rightarrow 2p\rightarrow 1s$.  We argued that the notion of an effective two-photon decay 
coefficient, useful and beloved for the $2s\rightarrow 1s$ decay of \HI, must be abandoned for the higher states in order to resolve these conceptual 
difficulties.  Rather one must resolve these two-photon decays as a function of frequency of the emitted photons and solve the full radiative transfer 
problem for two-photon emission and absorption.

In addition to the two-photon decays from the highly excited states, we have also considered Raman scattering and two-photon recombination.  These are 
physically and mathematically very similar processes, with the differences being moving one photon from the final to initial state (for Raman 
scattering) and moving the initial electron from a bound to a free state (for two-photon recombination).  Neither process has previously been 
investigated in the cosmological context.

Our major phenomenological findings are as follows:
\newcounter{conc}
\begin{list}{\arabic{conc}. }{\usecounter{conc}}
\item The traditional treatment of two-photon decays from $2s$ does not account for stimulated emission and re-absorption of the Ly$\alpha$ spectral 
distortion.  The maximal correction to $x_e$ from these effects is 0.6\% and over most of the redshift range the net effect is to delay recombination.  
We agree with both the physical picture and the calculation of past studies of these effects \cite{2006A&A...446...39C, 2006AstL...32..795K}.
\item Recombination is sped up by \subla\ two-photon decays from higher excited states; the most 
important initial state is $3d$.  Physically, inclusion of this effect involves a combination of two-photon emission and absorption which must be 
handled through a radiative transfer calculation; there is no relevant effective two-photon decay coefficient $\Lambda_{3d}$ analogous to that for 
$2s$.  We find a maximum change of 1.7\% in $x_e$.
\item \Supla\ two-photon decays initially speed up recombination by supplying another route to the ground 
state of hydrogen.  These photons produce a delayed feedback that slows down recombination at $z<1100$ as they redshift down to Ly$\alpha$.  The 
maximum effect on $x_e$ is 0.7\%.  The time dependence of the line profile plays a central role in this effect.
\item CMB photons can Raman scatter from hydrogen atoms in the $2s$ level and insert the atom into the ground state.  Again this initially speeds up 
recombination but produces photons blueward of Ly$\alpha$ that produce delayed feedback.  The maximum effect on $x_e$ is 1.0\%.
\item Direct two-photon recombinations to the ground state of hydrogen are negligible.
\item The net effect of the two-photon transitions on the CMB power spectrum is two-fold: (i) they suppress the temperature and polarization power 
spectra by up to 2\% at 
$\ell\sim 3000$ due to increased Silk damping from the faster early stages of recombination; and (ii) they increase the degree-scale polarization by 
$\sim$0.7\% due to the wider surface of last scatter.  At ``Fisher matrix'' level the net effect is $7\sigma$ for {\slshape Planck}.
\end{list}

We have constructed analytic approximations to all of the important two-photon processes that reproduce qualitatively and quantitatively the numerical 
radiative transfer results.  The error from using these analytic approximations instead of doing full two-photon radiative transfer will be $0.8\sigma$ 
for {\slshape Planck}.  This is reassuring: we anticipate that once {\em Planck} identifies the relevant portion of parameter space, this $\sim 
0.8\sigma$ correction to $x_e(z)$ can be added as a fudge factor. However much work lies ahead in order to make a full analysis of {\em Planck} data 
computationally feasible: our code even with the analytic approximations still takes $\sim 1$ day to run on a desktop machine, due primarily to the 
small step size of $\Delta\ln a=4.25\times 10^{-5}$.  We chose this time step because of our method of following feedback: the excitation in the 
$1s-np$ line requires us to look up the photon phase space density at a slightly earlier time on the red side of the $1s-(n+1)p$ line, and the time 
step must be small enough that the interpolation table for $f_{1s,(n+1)p-}$ has already been built. Iterative feedback methods, such as those used in 
Ref.~\cite{2007astro.ph..2143S}, can use much larger time steps and hence should be better suited to fast calculation.

In the future, we plan to incorporate these developments into a more complete hydrogen recombination model that would include effects not considered 
here.  Some of these effects, such as increasing the number of shells $n_{\rm max}$ and collisions, have previously received treatment in the 
literature \cite{2007MNRAS.374.1310C}, but ultimately one must {\em simultaneously} incorporate all important processes.  Another process that may 
interact with some of the effects here is Ly$\alpha$ diffusion and recoil, for which one would have to add Fokker-Planck terms to the radiative 
transfer equations described herein.  There is some work on this problem \cite{1989ApJ...338..594K, 1990ApJ...353...21K, 1994ApJ...427..603R, 
2008arXiv0801.3347G} and for the analogous problem in the \HeI\ 594$\,$\AA\ line \cite{2007astro.ph..2143S}; however given the importance of two-photon 
transitions, their optically thick nature, and the $>1$\% correction from effects that owe their existence to the time dependence of the line profile, 
it is clear that we need to solve the full time-dependent radiative problem including both Ly$\alpha$ diffusion and two-photon opacity.  This will be 
the subject of future work.  A final goal will be to speed up the recombination calculation to produce a recombination code suitable for inclusion into 
Markov Chain Monte Carlo codes for cosmological parameter estimation.

\begin{acknowledgments}
C.H. thanks Eric Switzer for a careful reading of the draft of this paper; and Jens Chluba, Rashid Sunyaev, Dan Grin, and Ulrich Jentschura for 
illuminating discussions about the physics.  This paper was improved by the suggestions from the referee, Douglas Scott.  C.H. is supported by DoE 
DE-FG03-92-ER40701.
\end{acknowledgments}

\appendix

\section{Matter temperature}
\label{app:matter}

This appendix constructs a solution for the matter temperature evolution valid at early times.
We can write a formal integral solution to Eq.~(\ref{eq:tmdot}) since the equation is (if we fix $x_e$) a first-order linear 
ODE in $T_{\rm m}/T_{\rm r}$.  The usual procedure gives
\begin{equation}
\frac{T_{\rm m}}{T_{\rm r}}(t) = -\int_0^t
\frac {J(\tilde t)}{1+J(\tilde t)}
e^{-K(\tilde t)} K'(\tilde t) d\tilde t,
\label{eq:tmdot2}
\end{equation}
where 
\begin{equation}
J = \frac{8x_e\sigma_{\rm T}a_{\rm r}T_{\rm r}^4}{3(1+f_{\rm He}+x_e)m_ecH}
\end{equation}
and $K$ is the integral defined by
\begin{equation}
K(\tilde t) = \int_{\tilde t}^t H(\bar t)[1+J(\bar t)] d\bar t,
\end{equation}
with $K(t)=0$ and $K'(\tilde t)<0$.  The initial value or ``particular solution'' is unimportant since $K(0)=\infty$.  Since $\int_0^te^{-K(\tilde 
t)} K'(\tilde t) d\tilde t=-1$, we may rewrite Eq.~(\ref{eq:tmdot2}) as
\begin{equation}
\frac{T_{\rm m}}{T_{\rm r}}(t) = 1 + \int_0^t
\frac 1{1+J(\tilde t)}
e^{-K(\tilde t)} K'(\tilde t) d\tilde t.
\label{eq:tmdot3}
\end{equation}
This is still only a formal solution because we cannot in general solve for $x_e$ without knowing $T_{\rm m}$.  However in the case $J\gg 1$
which occurs during recombination (but is violated at $z\lsim 500$), then the exponential factor $e^{-K(\tilde t)}$ kills the integrand in 
Eq.~(\ref{eq:tmdot3}) unless $\tilde t\approx t$.  In this case we may substitute $J(\tilde t)\approx J(t)$, and for $J\gg1$ we get
\begin{equation}
\frac{T_{\rm m}}{T_{\rm r}}(t) = 1 - \frac 1{J(t)},
\end{equation}
i.e. Eq.~(\ref{eq:tmdot4}).

\section{Two-photon transition rates}
\label{app:rates}

This appendix is concerned with calculating the transition rates for spontaneous two-photon decay ($nl\rightarrow 1s$), Raman scattering 
($nl\rightarrow 1s$), and spontaneous two-photon recombination to the ground state of hydrogen.  Stimulated rates and reverse reactions ($1s\rightarrow 
nl$ and two-photon ionization) are obtainable by insertion of the usual $1+f_\nu$ factor and by the principle of detailed balance, respectively.  Note 
that the two-photon selection rules allow these reactions only when the initial state has $l=0$ or $l=2$ (see DG05).  This also holds for two-photon 
recombination although in this case it is acceptable to sum the two cases because for the H$^++e^-$ continuum the $l$-sublevels are statistically 
populated.

We will have to consider both discrete and continuous states here.  The latter appear in summations over intermediate states in two-photon decay and 
Raman scattering, and as initial states in two-photon recombination.  They are characterized by an imaginary principal quantum number to agree with the 
usual formula, $E_e=-{\cal R}/n^2$ (we take the $\Im n>0$ branch), or alternatively by their wave number
\begin{equation}
k = \frac{\sqrt{2\mu E_e}}\hbar = i\frac{\sqrt\mu}{\hbar n}.
\end{equation}
The conventional normalization of the unbound states is that the radial
wavefunction $R(r)$ [with $r\psi(r,\theta,\phi) = R(r)Y_{lm}(\theta,\phi)$] is chosen to oscillate between $-1$ and $+1$, i.e. we normalize to a sphere
of radius $X=2$.  In this case, continuum states are formally separated in wave number by $\Delta k = \pi/X=\pi/2$.  Therefore in all calculations of
summations over intermediate $p$-states states, one should in actual computation make the replacement
\begin{equation}
\sum_{{\rm all}\,n} \rightarrow
\sum_{n=2}^\infty + \frac2\pi\int_0^\infty dk,
\end{equation}
where we will write ``all $n$'' for emphasis, and $n=ik/(\mu^{1/2}\hbar)$ for continuum states.

\subsection{Two-photon decay}

The spontaneous two-photon decay rate for the process
\begin{equation}
{\rm H}(nl) \rightarrow {\rm H}(1s) + h\nu + h\nu'
\end{equation}
is given by
\begin{equation}
\frac{d\Lambda}{d\nu} =
\frac{\alpha_{\rm fs}^6\nu^3\nu'{^3}}{108(2l+1){\cal R}^6}|{\cal M}|^2,
\label{eq:lambda}
\end{equation}
where $\alpha_{\rm fs}\approx 1/137$ is the fine structure constant, ${\cal R}$ is the hydrogen Rydberg constant, and the matrix 
element is
\begin{eqnarray}
{\cal M} &=& \frac 2{a_0^2}\sum_{{\rm all}\,N} \langle nl||{\bf r}||Np\rangle\langle Np||{\bf r}||1s\rangle
\nonumber \\ && \times
\Bigl(
\frac 1{1-N^{-2}-\nu/{\cal R}}
+\frac 1{1-N^{-2}-\nu'/{\cal R}}
\Bigr).\;\;\;\;
\label{eq:m}
\end{eqnarray}
(See Eq.~13 of Ref.~\cite{2007astro.ph..2144H}.)  The summation over intermediate states is carried out over both the discrete values of $N$ and the
continuous spectrum ($N^{-2}<0$, restricted of course to $p$ symmetry).  Note that ${\cal M}$ as defined here is dimensionless, and hence so is 
$d\Lambda/d\nu$.

One can write the total two-photon rate:
\begin{equation}
\Lambda_{nl} = \int_{(1-n^{-2}){\cal R}/2}^{(1-n^{-2}){\cal R}} \frac{d\Lambda}{d\nu}\,d\nu.
\label{eq:lnl}
\end{equation}
The cutoff at half the maximum frequency avoids double counting, since one photon is always above this cut and the other below.

\subsection{Raman scattering}

Now we are interested in the reaction
\begin{equation}
{\rm H}(nl) + h\nu' \rightarrow {\rm H}(1s) + h\nu.
\label{eq:raman}
\end{equation}
This reaction is related to the two-photon decay rate by crossing symmetry.  Therefore, whereas the two-photon decay rate was
\begin{equation}
\frac{d\Gamma_{nl,1s}}{d\nu}(2\gamma) = 
\frac{\alpha_{\rm fs}^6\nu^3\nu'{^3}}{108(2l+1){\cal R}^6}|{\cal M}|^2(1+f_\nu)(1+f_{\nu'}),
\end{equation}
the Raman scattering rate is
\begin{equation}
\frac{d\Gamma_{nl,1s}}{d\nu}({\rm Raman}) =
\frac{\alpha_{\rm fs}^6\nu^3\nu'{^3}}{108(2l+1){\cal R}^6}|{\cal M}|^2(1+f_\nu)f_{\nu'},
\end{equation}
where the matrix element ${\cal M}$ is given by Eq.~(\ref{eq:m}).
It is convenient for us to define the coefficient
\begin{equation}
\frac{dK_{nl}}{d\nu} = \frac{\alpha_{\rm fs}^6\nu^3\nu'{^3}}{108(2l+1){\cal R}^6}|{\cal M}|^2
\label{eq:k}
\end{equation}
that contains the matrix element and kinematic factors but not the state of the radiation field.  Then
\begin{equation}
\frac{d\Gamma_{nl,1s}}{d\nu}({\rm Raman}) =
\frac{dK_{nl}}{d\nu}(1+f_\nu)f_{\nu'}.
\end{equation}

\subsection{Two-photon recombination}

We also require the rate coefficient for two-photon recombination,
\begin{equation}
{\rm H}^+ + e^- \rightarrow {\rm H}(1s) + h\nu + h\nu',
\end{equation}
where the initial-state electron has kinetic energy $E_e$, and the final-state photons satisfy $h\nu+h\nu'=E_e-E_{1s}$.  We define the coefficient
$\alpha^{(2)}(\nu,\nu')n_en_p$ as the number of recombinations per unit time per unit frequency ($d\nu$) in a gas of monoenergetic elecrons and 
stationary protons, and in the absence of a stimulating radiation field.  The computation of this rate is similar to that of the two-photon decay, 
except that now the initial state is in the continuum ($n^2<0$).  This coefficient is a modification of Eq.~(\ref{eq:lambda}):
\begin{equation}
\alpha^{(2)}(\nu,\nu') = \sum_{l=0,2} \bar N_e(E,l)
\frac{\alpha_{\rm fs}^6\nu^3\nu'{^3}}{108(2l+1){\cal R}^6}|{\cal M}|^2,
\end{equation}
where $\bar N_e(E,l)$ is the mean number of electrons per normalization volume with angular momentum $l$, assuming unit electron density.  This is 
given by
\begin{equation}
\bar N_e(E,l) = 2\pi (2l+1) \frac X{k^2} = \pi(2l+1)\frac {\hbar^2X}{\mu E_e},
\end{equation}
where $X$ is the radius of the normalization sphere.  (This is most easily proved by 
considering the decomposition of a plane wave with unit probability density, $e^{i{\bf k}\cdot{\bf r}}$ into spherical harmonics, 
integrating the $l$th partial wave out to radius $X$, and taking the large-$X$ limit $\int_0^X r^2|j_l(kr)|^2dr\rightarrow X/2k^2$.)  This 
normalization volume cancels the $X^{-1/2}$ dependence of the matrix element ${\cal M}$ from the initial-state wave function.  For our choice of 
normalization, $X=2$ and we have
\begin{equation}
\alpha^{(2)}(\nu,\nu') = \sum_{l=0,2}
\frac{\pi\alpha_{\rm fs}^6\hbar^2\nu^3\nu'{^3}}{54{\cal R}^6\mu E_e}|{\cal M}|^2.
\label{eq:alpha2}
\end{equation}
The matrix element ${\cal M}$ is given by Eq.~(\ref{eq:m}).  Once again we restrict $\nu>\nu'$.
Note that the normalization of the free state forces $\psi({\bf r})$ to have units of cm$^{-1}$ instead of cm$^{-3/2}$, so for two-photon recombination 
${\cal M}$ has units of cm$^{1/2}$.  Thus $\alpha^{(2)}(\nu,\nu')$ has units of cm$^3$.

\subsection{Computation of matrix elements}

The two-photon matrix elements ${\cal M}$ can in principle be computed by brute force from Eq.~(\ref{eq:m}).  This was the approach taken in CS08 for 
hydrogen and in Hirata \& Switzer \cite{2007astro.ph..2144H} for helium.  Due to the cancellations among the various intermediate states for large $n$, 
however, we have chosen instead to use the Green function method, which is expected to be numerically stable even for very large $n$.  The Green 
function method is described in the recombination context by Eqs.~(24--26) of Ref.~\cite{2007astro.ph..2144H}, and more generally by 
Refs.~\cite{1928PNAS...14..253P, 1998JPhB...31.3743M}.

\bibliography{h2photon}

\end{document}